\documentclass[twocolumn,showpacs,preprintnumbers,amsmath,amssymb,prb,superscriptaddress]{revtex4}

\usepackage{graphicx}
\usepackage{latexsym}
\usepackage{dcolumn}
\usepackage{bm}
\usepackage{amsmath}

\begin{document}

\title{ Molecular orbital calculations of two-electron states for P donor solid-state spin qubits}

\author{L.M. Kettle} 
\affiliation{Department of Physics, National Taiwan University, Taipei 106, Taiwan (ROC). }
\author{Hsi-Sheng Goan} \email{goan@phys.ntu.edu.tw}
\affiliation{Department of Physics, National Taiwan University, Taipei 106, Taiwan (ROC). }
\author{Sean C. Smith}
\affiliation{Centre for Computational Molecular Science, University of Queensland, Brisbane Queensland 4072 Australia.}
\date{\today}

\begin{abstract}
We theoretically study the Hilbert space structure of two neighbouring P donor electrons in silicon-based quantum computer architectures. To use electron spins as qubits, a crucial condition is the isolation of the electron spins from their environment, including the electronic orbital degrees of freedom. We provide detailed electronic structure calculations of both the single donor electron wave function and the two-electron pair wave function. We adopted a molecular orbital method for the two-electron problem, forming a basis with the calculated single donor electron orbitals. Our two-electron basis contains many singlet and triplet orbital excited states, in addition to the two simple ground state singlet and triplet orbitals usually used in the Heitler-London approximation to describe the two-electron donor pair wave function. We determined the excitation spectrum of the two-donor system, and study its dependence on strain, lattice position and inter donor separation. This allows us to determine how isolated the ground state singlet and triplet orbitals are from the rest of the excited state Hilbert space. In addition to calculating the energy spectrum, we are also able to evaluate the exchange coupling between the two donor electrons, and the double occupancy probability that both electrons will reside on the same P donor. These two quantities are very important for logical operations in solid-state quantum computing devices, as a large exchange coupling achieves faster gating times, whilst the magnitude of the double occupancy probability can affect the error rate.
\end{abstract}


\maketitle

\section{\label{sec:one} Introduction}
Recently several designs for silicon-based quantum computer architectures have been proposed.\cite{kane,vrijen,skinner, ladd,sousa, friesen,lloyd2} In this work we concentrate our efforts on the Kane model,\cite{kane} which exploits a qubit array of nuclear spins of $^{31}$P dopants embedded within a silicon crystal matrix. The model is based on the use of $^{31}$P nuclear spins as qubits, with the donor electrons functioning to mediate control of single qubit operations via the hyperfine interaction, and interaction between individual qubits via the exchange interaction, and permit read-out of nuclear spin states. 

Performing logical operations on either electron-spin or nuclear-spin solid-state qubits requires precise control over single and two-qubit unitary operations, which corresponds to precise control over the electron-electron exchange interaction and the electron-nucleus hyperfine interaction in the Kane quantum computer. Here we calculate the exchange interaction as a function of the two donors' relative positions in the lattice and strain. We use multi-valley effective mass theory to calculate the single donor electron wave functions, these single donor orbitals combine to form our two-electron basis. This theory incorporates the Si crystal lattice effects by including the Si crystal Bloch functions into our single donor electron basis. Instead of using the Heitler-London (H-L) approximation, which has been used extensively in the literature so far for impurities in Si,\cite{me2, cam,cam2, koiller, koiller2} we describe our two-donor system using a rigorous molecular orbital method which employs our multi-valley single donor orbitals to form our two-electron basis, to calculate the exchange coupling more accurately.

An important feature necessary for quantum computing is to have well-characterized qubits, and for the two qubit case this is the ground state singlet and triplet two-electron states. It is meaningful to study the degree of proximity these targeted ground state orbitals are to the rest of the unwanted excited state Hilbert space.\cite{hu} This energy separation gives us an estimate for the conditions under which adiabaticity can be attained. We have pursued this goal using a molecular orbital method, which enables us to calculate a large number of two-electron energy levels (144), in the energy spectrum for our two donor system.

We used a molecular orbital method which includes the single donor ground state and first five excited states at each donor to form the two-electron basis. This yields a basis of 78 singlet states and 66 triplet states. This method not only gives us the exchange coupling, which is the difference between the ground singlet and triplet two-electron states, but also the spectrum of energy levels for the two donor system. For comparison we also calculated the H-L exchange coupling using just the symmetrized and anti-symmetrized products of the single donor ground states, and the Hund-Mulliken (H-M) exchange coupling which in addition to the H-L states, includes the two doubly occupied single donor ground states at each donor, in our two-electron basis. We calculated the exchange coupling and energy spectrum for our two-electron system as a function of donor position and strain. In addition, we also calculated the probability that both electrons will be on the same donor. This is also an important parameter for quantum gate operations, as these doubly occupied states can become a potential source of error. Several authors have similarly studied the exchange coupling, double-occupancy errors and adiabaticity of spin qubits in a number of different solid-state quantum computing architectures.\cite{hu,hu2,schliemann,barrett}

Much attention has been devoted to modeling the hyperfine and exchange interactions in these devices.\cite{me, me2, cam, cam2, koiller,koiller2, larionov, lloyd, smit,fang} Inter valley interference between degenerate conduction band minima in Si has been shown to lead to oscillations in the exchange coupling as a function of the donor pair positioning in the lattice.\cite{cam, cam2, koiller,koiller2} This poses serious problems for the fabrication of these devices, and leads to an extreme sensitivity of the exchange energy on the relative orientation of the P atoms. Koiller \emph{et al.}\cite{koiller} demonstrated that the introduction of external strain on the Si lattice partially lifts the valley degeneracy in the bulk Si. They showed that the inter valley effects could be reduced in some cases depending on the relative orientations of the donor pairs, whilst in other cases the donor exchange coupling remains oscillatory. The molecular orbital method we employ not only improves the calculation of the two donor electron wave functions, but we also use a more flexible basis than previous studies\cite{cam,cam2,koiller,koiller2} to calculate the single donor electron wave functions, which are used to construct the two-electron basis. 

\section{\label{sec:two} Quantum chemical models}

We advance beyond the simple H-L model for the two-electron wave function in the Kane device that has been previously considered.\cite{me2,cam, cam2,koiller,koiller2} In the molecular orbital method we use the single donor wave functions to form our basis states, and solve the 6-D Schr$\ddot{\mbox{o}}$dinger equation for the two electrons through a direct matrix diagonalisation.\cite{hu}

In the simplest case, the H-L approximation, the donor pair wave function is modeled as the symmetrized and anti-symmetrized products of the two single donor ground state wave functions (``$A_1$'' states) at each P nucleus, to form our singlet and triplet states respectively. In the H-M approximation, in addition to the two H-L states, the H-M basis incorporates the two ``ionised'' or ``polarised'' doubly occupied ground states, at each donor.

For the molecular orbital calculation we extended these bases to also include the first five excited states for each donor in our basis, in addition to the single donor ground states. This was chosen so that our basis included the six symmetry ground states for the single P donor, ($A_1, T_2$ and $E$ states). We performed calculations for the exchange coupling to see the effect that this larger basis has on lowering the energy of both the singlet and triplet ground states, and to improve upon and test the validity of using H-L theory to model the two-donor system over a range of device parameters.

We get a basis for our two-electron system which consists of the spatially symmetric singlet states, and anti-symmetric triplet states. Because the spin part of the singlet and triplet states are orthogonal, we can consider the singlet and triplet bases independently. Using six single donor orbitals on both qubits, we can form 78 singlet states and 66 triplet states.  The molecular orbital method has advantages over some quantum chemical methods as it includes the correlation between the two electrons, by virtue of including many two-electron orbitals to minimise the energy of the system.

\begin{figure} [htp]
\begin{center}

\setlength{\unitlength}{0.15cm}
\begin{picture}(60,25)

\put(10,0){\makebox(0,0){\bfseries {$Q_1$}  }}
\put(50,0){\makebox(0,0){\bfseries {$Q_2$}  }}

\put(10.5,5){\vector(1,0){10}}

\put(10.5,5){\vector(0,1) {10}}

\put(7.5, 13){\makebox(0,0) {$z$   }}

\put(19,2){\makebox(0,0){$y$}}

\put(10.5,5){\vector(1,0) {40}}

\put(42,2){\makebox(0,0){$\mathbf{R}$}}

\put(10.5,5){\vector(1,1) {13.33}}

\put(50.5,5){\line(-2,1) {13.33}}

\put(37.17,11.665){\vector(-2,1) {13.33}}

\put(50.5,5){\vector(-1,1) {13.33}}

\put(10.5,5){\line(2,1) {13.33}}

\put(23.83,11.665){\vector(2,1) {13.33}}

\put(23.83,18.33){\vector(1,0) {13.34}}

\put(37.17,18.33){\circle*{1}}

\put(23.83,18.33){\circle*{1}}

\put(10.5,5){\circle{2}}
\put(50.5,5){\circle{2}}

\put(21,20){\makebox(0,0){$e_1$}}

\put(40,20){\makebox(0,0){$e_2$}}

\put(18,17){\makebox(0,0){$\mathbf{r_1}$}}

\put(24,9){\makebox(0,0){$\mathbf{r_2}$}}

\put(34,10){\makebox(0,0){$\mathbf{r_1-R}$}}

\put(44,17){\makebox(0,0){$\mathbf{r_2-R}$}}

\put(30,20.5){\makebox(0,0){$\mathbf{r_2-r_1}$}}

\end{picture}
\end{center}

\caption{\label{fig:fig1} Co-ordinate geometry of our two-electron problem.}

\end{figure}
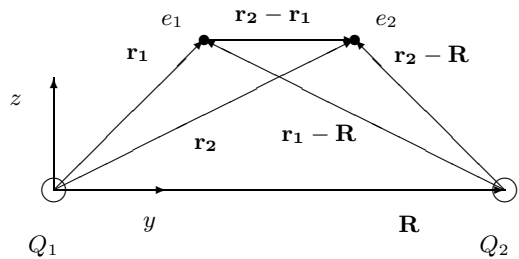

We show the geometry of our two-electron problem in Fig.~\ref{fig:fig1} for two P donors, $Q_1$ and $Q_2$, embedded in the Si lattice, (the origin is at $Q_1$). For the singlet (symmetric spatial orbitals) basis we form the following two-electron wave functions, from our basis of single donor orbitals, $\Psi_{Q_1}^{e_n}(\bm{r})$ and $ \Psi_{Q_2}^{e_m}(\bm{r-R})$,  ($e_n$th state at $Q_1$, and $e_m$th state at $Q_2$ respectively):
\begin{widetext}
\begin{eqnarray*}
\Psi^S_{1-21} &=& \frac{1}{\sqrt{2(1+\delta_{nm})}} \left[ \Psi_{Q_1}^{e_n}(\bm{r_1}) \Psi_{Q_1}^{e_m}(\bm{r_2}) + \Psi_{Q_1}^{e_n}(\bm{r_2}) \Psi_{Q_1}^{e_m}(\bm{r_1}) \right], \\
&& \mbox{for   } \quad n=0 \mbox{ to }5 \quad \mbox{and } \quad m = n\mbox{ to }5 . \\
\Psi^S_{22-42} &=& \frac{1}{\sqrt{2(1+\delta_{nm})}} \left[ \Psi_{Q_2}^{e_n}(\bm{r_1-R}) \Psi_{Q_2}^{e_m}(\bm{r_2-R}) + \Psi_{Q_2}^{e_n}(\bm{r_2-R}) \Psi_{Q_2}^{e_m}(\bm{r_1-R}) \right], \\
&& \mbox{for   } \quad n=0\mbox{ to }5 \quad \mbox{and } \quad m = n\mbox{ to }5 . \\
\Psi^S_{43-78} &=& \frac{1}{\sqrt{2(1+|S_{nm}|^2)}} \left[ \Psi_{Q_1}^{e_n}(\bm{r_1}) \Psi_{Q_2}^{e_m}(\bm{r_2-R}) + \Psi_{Q_1}^{e_n}(\bm{r_2}) \Psi_{Q_2}^{e_m}(\bm{r_1-R}) \right], \\
&& \mbox{for   } \quad n= 0\mbox{ to }5 \quad \mbox{and } \quad m = 0\mbox{ to }5, 
\end{eqnarray*}
\begin{eqnarray*}
&& \mbox{where   } \quad S_{nm} = \int d\bm{r} \Psi_{Q_1}^{e_n}(\bm{r}) \Psi_{Q_2}^{e_m}(\bm{r-R}).
\end{eqnarray*}
\end{widetext}
Here we see that the two-electron singlet donor wave functions $\Psi^S_{1-21}$ and $\Psi^S_{22-42}$, are the doubly occupied singlet states located at $Q_1$ and $Q_2$ respectively. The two-electron states, $\Psi^S_{43-78}$, are the ``Heitler-London like'' singlet states formed from the single donor ground state and excited state wave functioins.

Similarly for the triplet (anti-symmetric spatial orbitals) basis we obtain the following two-electron wave functions:
\begin{widetext}
\begin{eqnarray*}
\Psi^T_{1-15} &=& \frac{1}{\sqrt{2}} \left[ \Psi_{Q_1}^{e_n}(\bm{r_1}) \Psi_{Q_1}^{e_m}(\bm{r_2}) - \Psi_{Q_1}^{e_n}(\bm{r_2}) \Psi_{Q_1}^{e_m}(\bm{r_1}) \right], \\
&& \mbox{for   } \quad n=0\mbox{ to }5 \quad \mbox{and } \quad m = n+1\mbox{ to }5, \quad (m\neq n). \\
\Psi^T_{16-30} &=& \frac{1}{\sqrt{2}} \left[ \Psi_{Q_2}^{e_n}(\bm{r_1-R}) \Psi_{Q_2}^{e_m}(\bm{r_2-R}) - \Psi_{Q_2}^{e_n}(\bm{r_2-R}) \Psi_{Q_2}^{e_m}(\bm{r_1-R}) \right], \\
&& \mbox{for   } \quad n=0\mbox{ to }5 \quad \mbox{and } \quad m = n+1\mbox{ to }5, \quad (m\neq n). \\
\Psi^T_{31-66} &=& \frac{1}{\sqrt{2(1-|S_{nm}|^2)}} \left[ \Psi_{Q_1}^{e_n}(\bm{r_1}) \Psi_{Q_2}^{e_m}(\bm{r_2-R}) - \Psi_{Q_1}^{e_n}(\bm{r_2}) \Psi_{Q_2}^{e_m}(\bm{r_1-R}) \right], \\
&& \mbox{for   } \quad n= 0\mbox{ to }5 \quad \mbox{and } \quad m = 0\mbox{ to }5.
\end{eqnarray*}
\end{widetext}

It is clear that the singlet and triplet bases contain the original H-L states, $\Psi^S_{43}$ in the singlet basis, and $\Psi^T_{31}$ in the triplet basis. For the H-M calculation we include the two additional ``ionised'' or doubly occupied ground states, $\Psi^S_1$ and $\Psi^S_{22}$ in our singlet basis. In the extended molecular orbital basis, we consider all 78 singlet states and 66 triplet states in our two donor electron Hamiltonians for the singlet and triplet bases respectively.

\section{\label{sec:three} Numerical method}
\subsection{Solution of the single donor wave function}

To obtain our two-electron states we first need to evaluate the single donor wave functions at each donor to use in the two-electron basis. We did this in the case of no strain and with uniaxially strained Si. We calculated the single donor orbitals using multi-valley effective mass theory. We use a basis for the multi-valley single donor wave function which includes the  full Bloch structure at each conduction band minimum, in our basis functions.

The Kohn-Luttinger\cite{kohn1} form of the wave function for a donor (where we define the P nucleus to be at the origin on a substituted Si atom site) is given by:
\begin{eqnarray}
\Psi(\bm{r}) &=& \sum _{\mu = 1}^{6} \alpha_\mu F^{(\mu)}(\mathbf{r}) \psi^{0}_{k_\mu}(\bm{r}) ,\nonumber \\
&=& \sum _{\mu = 1}^{6} \alpha_\mu F^{(\mu)}(\mathbf{r}) e^{i\bm{k_\mu}.\mathbf{r}} u_{k_\mu}(\mathbf{r + R_0}), \label{eq:s1}
\end{eqnarray}
where we choose $\bm{R_0} = (a^0/8)(1,1,1) = \bm{R^a_0}$,\cite{Yu,chelikowsky} as we take the origin to be at an substitutional P donor site, here $a^0 = 0.543$nm is the length of the unit cell. The term $\bm{R^a_0}$ arises because Si has two atoms per lattice point in the unit cell, where the Si atoms are chosen to be displaced a distance of $\pm \bm{R^a_0}$ away from each lattice point in the unit cell.\cite{Yu,chelikowsky} For substitutional donors $\bm{R_0} = \pm \bm{R^a_0}$.
 
$F^{(\mu)}(\bm{r})$ is the donor envelope function at the conduction band minimum, $\bm{k_\mu}$, and $\psi^{0}_{k_\mu}(\bm{r})$ is the silicon crystal Bloch function at the conduction band minimum, $\bm{k_\mu}$, where $\bm{k_{^1_2}} = \pm \bm{ k_z} = (0,0, \pm k) 2 \pi/a^0$ etc. and $k=0.85$.  

The multi-valley effective mass equation for a P donor in Si under strain is:\cite{pant1}
\begin{eqnarray}
\sum _{\mu = 1}^{6} \alpha_{\mu} e^{i(\mathbf{k_\mu} - \bm{k_\nu}).\mathbf{r}}[T_\mu (-i \nabla) + U(r) + H_{strain} - E] && \nonumber \\
 \times  F^{(\mu)}(\mathbf{r}) = 0, && \label{eq:s1a} 
\end{eqnarray}
where:
\begin{eqnarray}
T_1 (-i \nabla) &\equiv&  \left( \frac{\partial^2}{\partial x^2} + \frac{\partial^2}{\partial y^2} \right)   + \gamma \frac{\partial^2}{\partial z^2},  \nonumber \\
&=& T_2 (-i \nabla) , \nonumber \\
T_3 (-i \nabla) &\equiv&  \left( \frac{\partial^2}{\partial x^2} + \frac{\partial^2}{\partial z^2} \right)   + \gamma \frac{\partial^2}{\partial y^2},  \nonumber \\
&=& T_4 (-i \nabla) , \mbox{   etc.} \nonumber 
\end{eqnarray}
Here $T_\mu$ are the anisotropic kinetic energy terms, due to the anisotropy of the conduction band minima in Si. The impurity potential, $U(r)$, is the potential term due to the effective $+1$ charge of the P nucleus in the Si lattice. Here we model the impurity potential as a screened Coulombic potential, $U(r) =  2/r$. We are using atomic units, where the unit of length $[ a_B] =  \hbar^2 \epsilon / m_\perp e^{'2}= 31.667\dot{\mbox{A}} $ and unit of energy  $[ E_B] =  m_{\perp} e^{'4} / 2 \hbar^{2} \epsilon^2  = 19.9436$meV, where $\epsilon = 11.4$a.u. and $\gamma =  m_{\perp}/m_{\parallel} = 0.2079$.\cite{me} $H_{strain}$ is the potential due to uniaxial strain along the $z$-direction which we will define later.

We expanded the donor electron envelope wave function, $F^{(\mu)}$, in a basis of the single-valley zero field envelope wave functions, $F_j^{(\mu)}$, at each of the six conduction band minimum, $\mathbf{k_\mu}$. Here $F_j^{(\mu)}$ are the eigenfunctions of the single-valley zero field Hamiltonian, $H^{(\mu)}_0 = T_\mu (-i \nabla) + U(r)$. We have discussed previously\cite{me} how we obtained the single-valley zero-field wave functions, $F_j^{(\mu)}$, by expanding these single-valley wave functions in a basis of deformed hydrogenic orbitals.

In Eq. (\ref{eq:s1}) we expanded the donor electron wave function, $\Psi(\mathbf{r})$, in a basis of the donor electron envelope functions, $F^{(\mu)}$, at each minimum. In addition we can also expand $F^{(\mu)}$ in our basis of single-valley donor electron wave functions, $F_j^{(\mu)}$:
\begin{eqnarray}
\Psi(\mathbf{r}) &=& \sum_{\mu = 1}^{6} \alpha_\mu e^{i\mathbf{k_\mu}.\mathbf{r}} u_{k_\mu}(\mathbf{r + R_0}) F^{(\mu)}(\mathbf{r}), \nonumber \\
&=& \sum_{\mu = 1}^{6} e^{i\mathbf{k_\mu}.\mathbf{r}} u_{k_\mu}(\mathbf{r + R_0}) \sum_j C_j^{(\mu)} F_j^{(\mu)}(\mathbf{r}),\nonumber \\ \label{eq:s2} \\
\mbox{where  } F_j^{(\mu)}(\mathbf{r}) &=& \sum_{n,l,m} B^{(\mu)}_{nlm}\phi^{(\mu)}_{nlm}(x,y,z,a ,\beta), \nonumber
\end{eqnarray}
and $C_j^{(\mu)}$ are the expansion co-efficients for our basis functions $F_j^{(\mu)}(\bm{r})$. We see that the single-valley envelope functions, $F_j^{(\mu)}$, are in turn, also a sum of basis functions: the deformed hydrogenic orbitals, $\phi^{(\mu)}_{nlm}(x,y,z,a ,\beta)$, given already in a previous paper.\cite{me} The co-efficients $B^{(\mu)}_{nlm}$ are determined already since $F_j^{(\mu)}(\mathbf{r})$ are the eigenfunctions of the single-valley Hamiltonians, $H_0^{(\mu)}$, ie. $H_0^{(\mu)} F_j^{(\mu)}(\mathbf{r}) = E_j^0 F_j^{(\mu)}(\mathbf{r})$.

Note that including the expansion co-efficients $C_j^{(\mu)}$ in Eq.~(\ref{eq:s2}) is a generalisation of the calculations of Wellard \emph{et al.}\cite{cam,cam2} where we have removed the restriction that the donor wave function be composed of equal contributions from the six conduction band minima. Clearly this restriction breaks down when an external strain is applied as this will break the degeneracy of the six conduction band minima.

So now Eq.~(\ref{eq:s1a}) becomes:              
\begin{eqnarray}
\sum _{\mu = 1}^{6} e^{i(\mathbf{k_\mu}-\bm{k_\nu}).\mathbf{r}} \sum_j C_j^{(\mu)} [H_0^{(\mu)} + H_{strain}  - E] F_j^{(\mu)}(\mathbf{r}) &=& 0. \nonumber \\  \label{eq:s3}
\end{eqnarray}                                                                         
We now multiply Eq.~(\ref{eq:s3}) by $F_i^{*(\nu)}(\mathbf{r})$ and integrate over $\mathbf{r}$. The orthonormality of this basis is enforced by the $e^{i(\mathbf{k_\mu} - \mathbf{k_\nu}).\mathbf{r}}$ terms which appear in the matrix elements, and due to their rapidly oscillating nature average to zero unless $\mathbf{k_\mu} = \mathbf{k_\nu}$.\cite{cam} 

In the standard effective mass treatment, the inter valley mixing terms which couple the envelope functions at different conduction band minima in the above approximation are neglected, and six independent equations are obtained. For the higher donor excited states this is a valid approximation, as their energies agree quite well with calculations using only single valley effective mass theory.\cite{faulkner} However, we need to consider the inter valley coupling for the donor ground state, which has the effect of lifting the six-fold degeneracy of the $1S$ states predicted by the one-valley effective mass equations. In order to obtain the correct symmetry states for the donor ground state, of a singlet ($A_1$), a triplet ($T_2$) and doublet ($E$), we add empirically determined parameters to our Hamiltonian, as was done by Koiller \emph{et al.}\cite{koiller}

Hence the multi valley effective mass Eq. (\ref{eq:s3}) becomes:
\begin{eqnarray}
\lefteqn{ \delta_{1j} \sum_{\mu=1}^6 C_1^{(\mu)} \Delta^{\mu,\nu} +  \delta_{\mu\nu} \delta_{ij} C_i^{(\nu)} E_i^0 } \nonumber \\
& +& \delta_{\mu\nu} \sum_j C_j^{(\nu)} \int d\mathbf{r}   F_i^{*(\nu)}(\mathbf{r}) \left[ H_{strain} \right] F_j^{(\nu)} (\mathbf{r}) \nonumber \\ 
&=& E \delta_{\mu\nu} \delta_{ij} C_i^{(\nu)}.   \label{eq:s4}
\end{eqnarray}
where $\Delta^{\mu,\nu} = \left\{ \begin{array}{ll} 0, & \mbox{if }\mu = \nu, \\  -2.1934\mbox{meV}, & \mbox{if  }\mu, \nu \mbox{ are on perpendicular} \\ & \mbox{symmetry valleys,}  \\ -1.535\mbox{meV}, & \mbox{if  }\mu, \nu \mbox{  are on opposite} \\ & \mbox{symmetry valleys}. \end{array} \right. $
Here we also scale the single valley ground state energy, $E_1^0=-35.19$meV, to reproduce accurately the experimental splitting for the P donor ground state. We only scaled the single valley ground state energy because the single valley calculation reproduces the higher excited state energies reasonably accurately.\cite{faulkner}

When we consider the effect of strain on the single donor orbitals, we consider its effect only on the lowest six energy states, because later we construct our two-electron basis from these lowest six single donor states. Here we follow the treatment of Koiller \emph{et al.}\cite{koiller} and introduce the relative energy shifts due to uniaxial strain along the $z$-direction, in terms of a dimensionless valley strain parameter, $\chi$. In their paper they discuss the physical relevance and tuning of this parameter. For our purposes we consider four cases of the strain parameter, corresponding to $\chi = 0,-1,-5$ and $-20$. Negative values of $\chi$ correspond to tensile strain, which favours the $z$-envelopes energetically, and $\chi=-20$ represents the realistic situation of Si grown over relaxed Si$_{0.8}$Ge$_{0.2}$.\cite{koiller} 

To evaluate the $H_{strain}$ terms we first need to define $\mu = 1,2,3,4,5,6$ to correspond to the $z,-z,y,-y,x,-x$ valleys respectively. Now the $H_{strain}$ terms in Eq.~(\ref{eq:s4}) become:
\begin{eqnarray}
\lefteqn{ \delta_{1j} \sum_{\mu=1}^6 C_1^{(\mu)} \int d\mathbf{r}   F_1^{*(\mu)}(\mathbf{r}) \left[ H_{strain} \right] F_1^{(\mu)} (\mathbf{r}) } \nonumber \\
 & = & \delta_{1j} \left( \sum_{\mu=1}^2 C_1^{(\mu)} (2 \chi \Delta_C) + \sum_{\mu=3}^6 C_1^{(\mu)} (-\chi \Delta_C)\right), \label{eq:hs1}
\end{eqnarray}
where $\Delta_C = 2.16$meV is used to be consistent with Koiller \emph{et al.}\cite{koiller}

\begin{figure} [t!]
\begin{center}
\includegraphics[width=2.5in,angle=0]{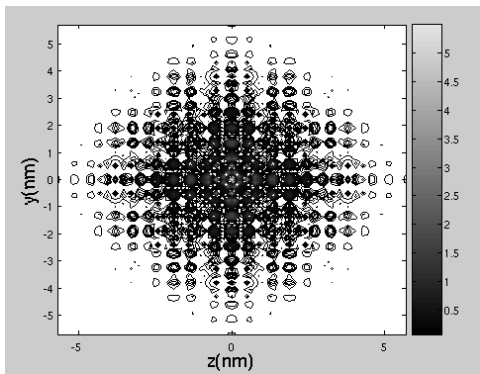}
\end{center}
 \caption{\label{fig:s1} Contour plot of the ground state electron density in the $yz$-plane for $A_1$ state without any strain applied. Here the P nucleus is located at the origin.}
\end{figure}
\begin{figure} [t!]
\begin{center}
\includegraphics[width=2.5in,angle=0]{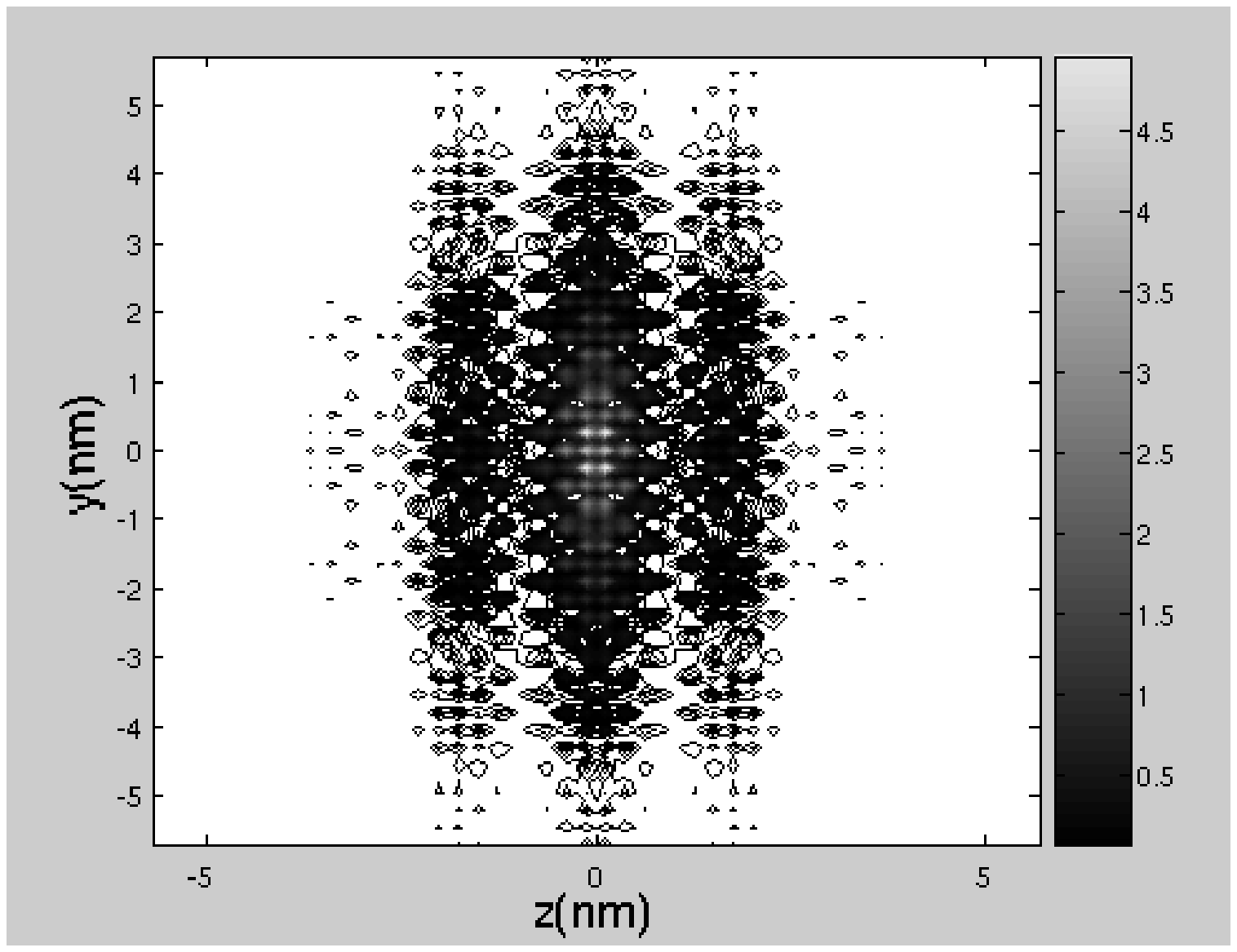}
\end{center}
 \caption{\label{fig:s2} Contour plot of the ground state electron density in the $yz$-plane with a strain parameter $\chi=-20$. Here the P nucleus is located at the origin.}
\end{figure}
\begin{figure} [t!]
\begin{center}
\includegraphics[width=2.5in,angle=0]{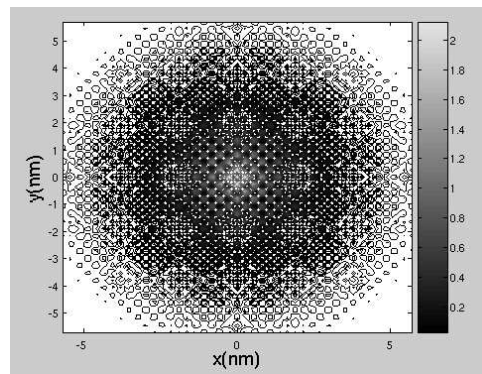}
\end{center}
 \caption{\label{fig:s3} Contour plot of the ground state electron density in the $xy$-plane with a strain parameter $\chi=-20$. Here the P nucleus is located at the origin.}
\end{figure}

We plot the ground state electron density without any external strain, $\chi = 0$, and with strain applied, $\chi = -20$, in Fig.~\ref{fig:s1} to \ref{fig:s3} using Eq.~(\ref{eq:s2}). Here the Bloch functions are obtained using the empirical pseudo-potential technique \cite{chelikowsky,cohen} and the P donor envelope functions are obtained from the multi-valley effective mass equations.  Figure~\ref{fig:s1} is a contour plot of the ground state electron density in the $yz$-plane for the symmetric $A_1$ state corresponding to zero strain, where the contribution from all six valleys are equivalent. However with a strain applied in the $z$-direction, we see  in Fig.~\ref{fig:s2}, where we plotted the electron density in the $yz$-plane, the effective Bohr radius in the $z$-direction is reduced. This is because with an external strain applied, the six-valley degeneracy of the symmetric $A_1$ ground state is broken. This can be seen from Eq.~(\ref{eq:hs1}), and the lowest energy state is the one in which the effective Bohr radius in the direction parallel to the strain is reduced, ie. the $F^{\pm z}_1(1S)$ states. In contrast in Fig.~\ref{fig:s3}, where we plotted the ground state density in the $xy$-plane, we see that the strain (applied in the $z$-direction) is equivalent in these two directions, and the effective Bohr radii has increased.

Table~\ref{table2} reports the energy splitting between the ground state and first excited state for a single donor electron for different magnitudes of strain applied. The energy levels become closer together when a strain is applied, and the ground ``$A_1$'' state is no longer degenerate in the six valleys, and we find that the $F_{\pm z}$ valleys become more favored. This leads to a smaller effective Bohr radius in the $z$-direction, and larger effective Bohr radii in the $x,y$-directions, which was demonstrated already in Fig.~\ref{fig:s2} and \ref{fig:s3}. 


\begin{table}[h!]\caption{\label{table2} Energy splitting between the ground state and the first excited state for a single donor electron.}
\vspace{0.1cm}
\begin{center}
\begin{tabular}{|c|c|} \hline
$\chi$ & $\Delta E$ (meV)\\ \hline
0 & 11.847 \\
-1 & 8.316 \\
-5 & 4.383 \\
-20 & 3.378 \\ \hline
\end{tabular}
\end{center}
\end{table}

\subsection{Solution of the two-electron donor pair wave function }

Once the single donor orbitals are known, we are then able to evaluate the 6-D two electron Hamiltonian matrices for both our singlet and triplet bases, $H_{2e}^S$ and $H_{2e}^T$, and the singlet and triplet overlap matrices, $S^S$ and $S^T$. Here the 6-D two electron Hamiltonian operator is:
\begin{widetext}
\begin{eqnarray*}
H_{2e} &=& -\nabla_{anis}^2(\mathbf{r_1}) -\nabla_{anis}^2(\mathbf{r_2}) - \frac{2}{|\mathbf{r_1}|} - \frac{2}{|\mathbf{r_2}|} - \frac{2}{|\mathbf{r_1 - R}|}- \frac{2}{|\mathbf{r_2 - R}|}  + \frac{2}{|\mathbf{r_1 - r_2}|} + H_{strain}(\mathbf{r_1}) + H_{strain}(\mathbf{r_2}) .
\end{eqnarray*}
\end{widetext}
Thus we need to evaluate both the singlet and triplet Hamiltonian matrix elements, $\langle  \Psi^S_{1-78} | H_{2e}^S | \Psi^S_{1-78} \rangle$ and $\langle  \Psi^T_{1-66} | H_{2e}^T | \Psi^T_{1-66} \rangle$ respectively, and singlet and triplet overlap matrix elements, $S(i,j) = \langle  \Psi_{i}  | \Psi_{j} \rangle$, for varying inter donor separation and strain. 

Since our basis functions $\Psi_{Q_1}^{e_n}(\bm{r})$ and $\Psi_{Q_2}^{e_m}(\bm{r - R})$ are eigenfunctions of the single electron Hamiltonian operator at each donor, we use this property in evaluating the two electron Hamiltonian matrix elements:
\begin{widetext}
\begin{eqnarray*}
\left[ -\nabla_{anis}^2(\mathbf{r}) - \frac{2}{\bm{r}} + H_{strain}(\bm{r})  \right] \Psi_{Q_1}^{e_n}(\bm{r}) &=& E^{e_n}_{Q_1} \Psi_{Q_1}^{e_n}(\bm{r}) , \\
\left[ -\nabla_{anis}^2(\mathbf{r}) - \frac{2}{\bm{r-R}} + H_{strain}(\bm{r}) \right] \Psi_{Q_2}^{e_m}(\bm{r-R}) &=& E^{e_m}_{Q_2} \Psi_{Q_2}^{e_m}(\bm{r-R}).
\end{eqnarray*}
\end{widetext}

Here when we calculate the matrix elements for the singlet and triplet Hamiltonian and overlap matrices, we retain the single plane wave part, $e^{i\bm{k_\mu}.\mathbf{r}}$, of the Bloch functions at each minima in the expansion for the single donor electron wave functions, in the integrands involving these wave functions. This leads to the inherent oscillations in the exchange energy due to the inter valley interference between these terms at the degenerate conduction band minima. We still neglect the periodic part of the Bloch functions, $u_{k_\mu}(\mathbf{r}+\bm{R_0})$, in the integrands. It has been shown\cite{cam} that this is an excellent approximation, and it was impossible to distinguish between the results for the exchange coupling using this approximation, and those including the detailed Bloch structure. We did this to make the calculations more tractable over a larger range of device parameters.

Once we derived the matrix elements for both the singlet and triplet Hamiltonian and overlap matrices, we needed to solve a generalised eigenvalue problem for both the singlet and triplet case. This is because the two-electron states are not necessarily orthogonal, since the single electron wave functions, $\Psi_{Q_1}^{e_n}(\bm{r})$ and $ \Psi_{Q_2}^{e_m}(\bm{r-R})$, are not orthogonal. We have:
\begin{equation}
 H_{2e} \bm{c} = E S \bm{c}. \label{eq:1}
\end{equation}
Here $\bm{c}$ is a vector of the coefficients of the two-electron basis functions. To solve this we first need to compute the Cholesky factorisation for the overlap matrix $S$, to give $S=LL^+$. We did this using a standard numerical subroutine. Once we had obtained the Cholesky factorisation, we used this to transform Eq.~(\ref{eq:1}) into the standard eigenvalue problem using another subroutine:
\begin{equation}
[ L^{-1} H_{2e} (L^+)^{-1} ] [L^+ \bm{c}] =  E [L^+ \bm{c}] .\label{eq:2}
\end{equation}
Once we derived the standard eigenvalue problem, we used a standard eigenvalue solver, to diagonalise Eq.~(\ref{eq:2})  to obtain the energies $E$, for the singlet and triplet states.

The most computationally expensive task in our molecular orbital calculations is the computation of the 6-D two-electron integrals in the singlet and triplet Hamiltonian matrix elements. In the singlet basis this required 3081 6-D integrals to be performed, and for the triplet basis 2211 6-D integrals to be performed. We have reduced this task greatly by only calculating the identical 6-D integrals in both the singlet and triplet bases once. This means calculating 4131 6-D integrals in total. This is also a better numerical practice as it means that when we calculate the energy splitting between the ground singlet and triplet states, (which can be very small), we are using the same integral evaluations to calculate both quantities, $E_T$ and $E_S$. Thus the exchange energy calculated $J=E_T - E_S$ will be more accurate, as the same numerical errors will be involved in both quantities.

We have also increased the efficiency and speed of our code by modifying a standard Monte Carlo subroutine used to numerically evaluate the 3-D and 6-D integrals. We did this because the integrals all require evaluations of our single donor basis functions, $\Psi_{Q_1}^{e_n}(\bm{r})$ and $\Psi_{Q_2} ^{e_m}(\bm{r-R})$. We have greatly reduced the complexity and computing time for these calculations by evaluating these common basis functions on a grid, before inputting these functions into the Monte Carlo subroutine. This has provided a speed-up of our calculations of the order of 100 times.

\section{\label{sec:four} Results and discussion }
\subsection{Results using full molecular orbital calculation}

\begin{figure} [t!]
\begin{center}
 \includegraphics[width=2in,angle=270]{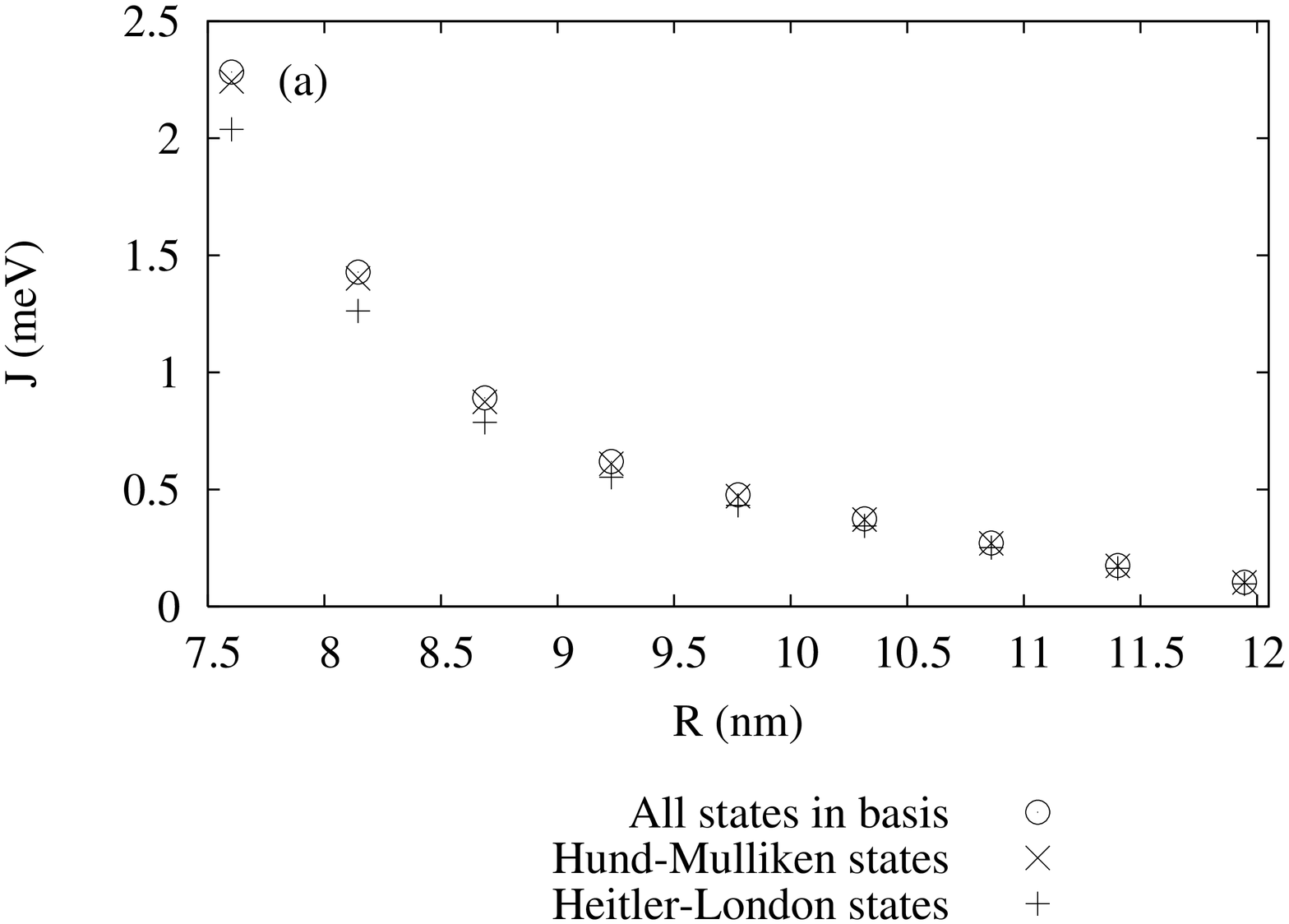} 
 \includegraphics[width=2in,angle=270]{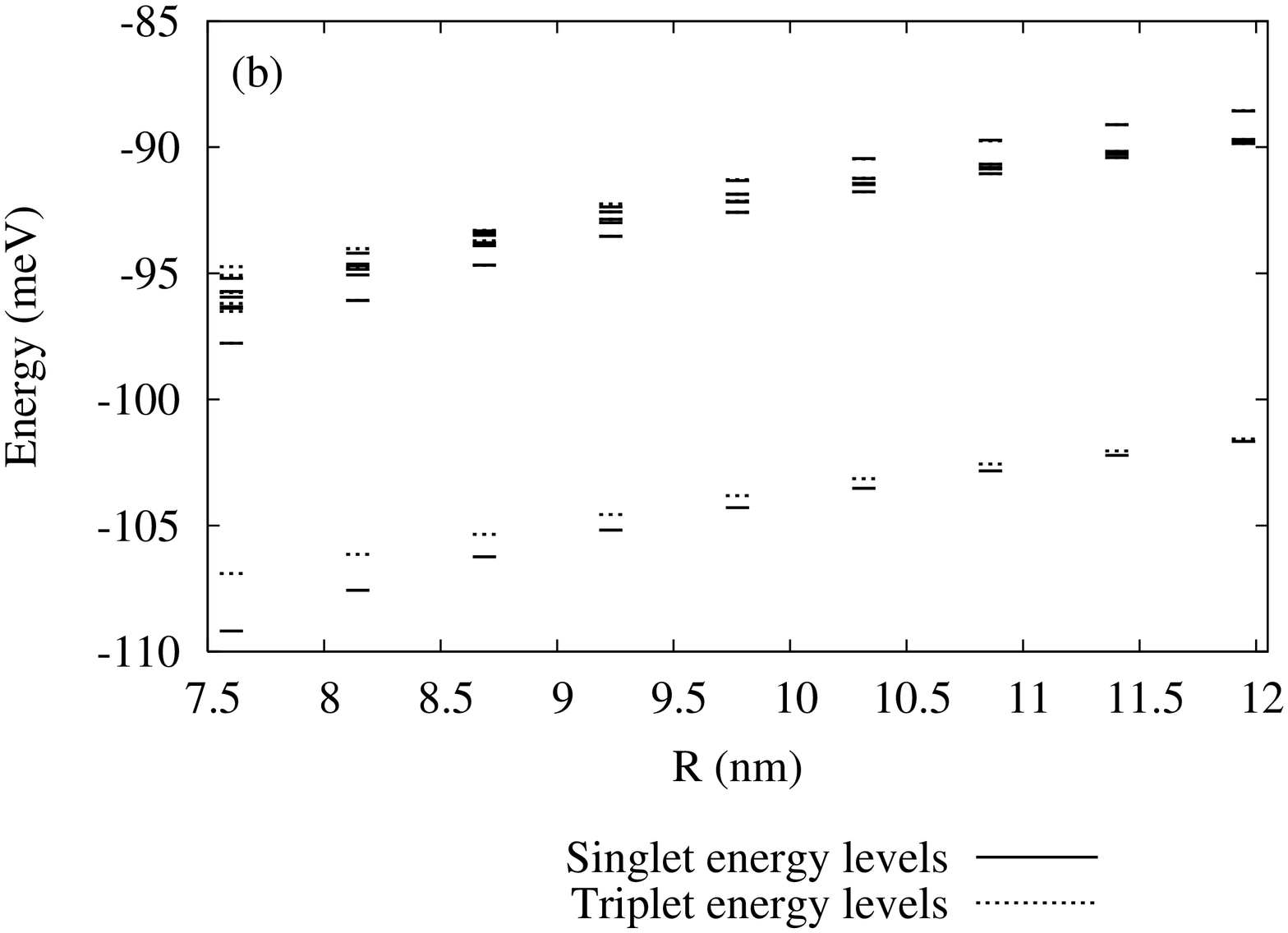} 
\end{center}
 \caption{\label{fig:fign1a} We compare the exchange coupling at lattice sites along the $[010]$ or $y$ direction for small magnitudes of $\bm{R}$, calculated using our three quantum chemical models in (a): using the H-L states, H-M states and our extended molecular orbital basis, for zero strain. In (b) we plot the singlet and triplet two-electron energy levels using our extended basis. Here we only consider values of $\bm{R}$ such that both P donors are on substitutional donor sites.}
\end{figure}

\begin{figure} [t!]
\begin{center}
 \includegraphics[width=2in,angle=270]{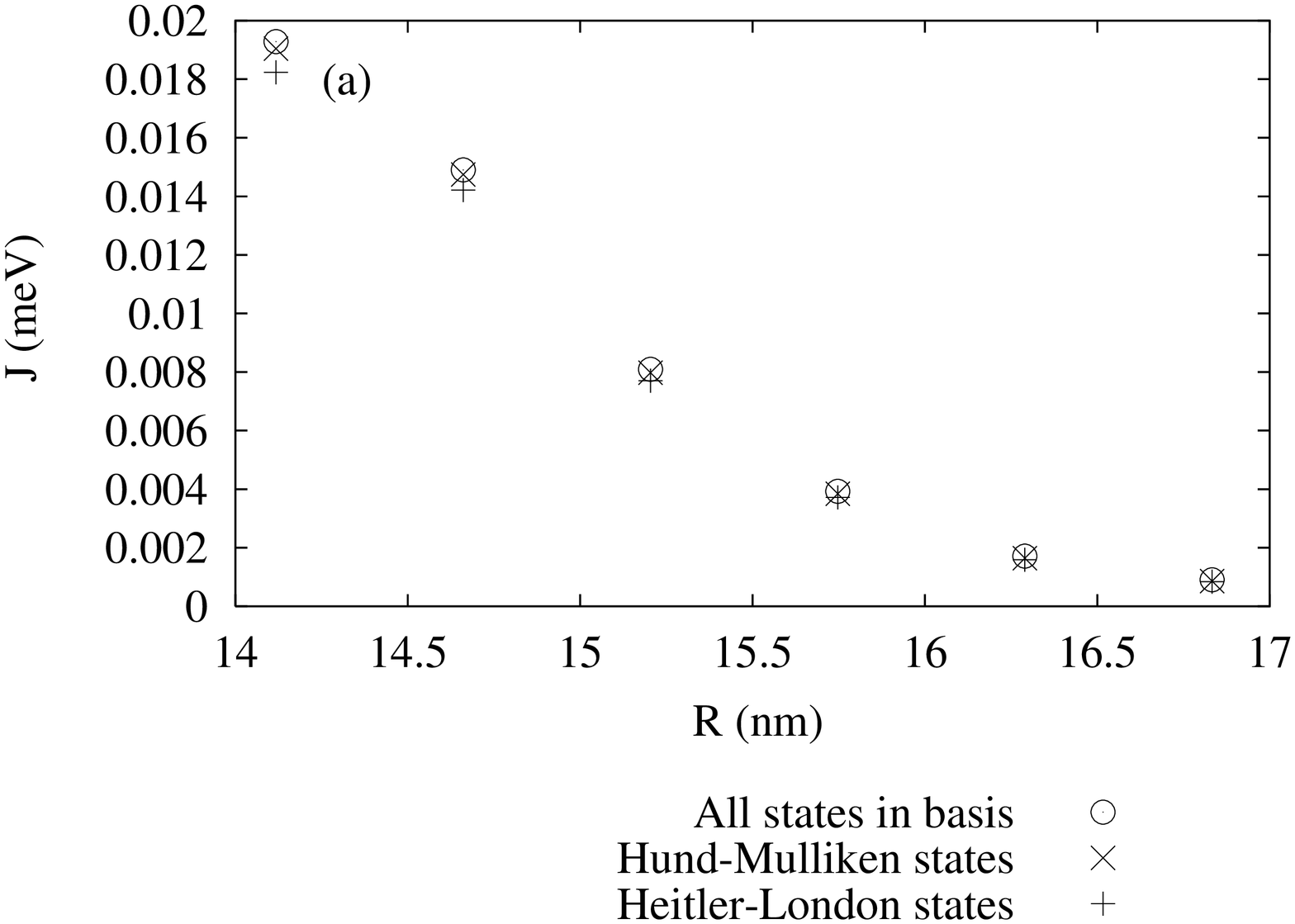} 
 \includegraphics[width=2in,angle=270]{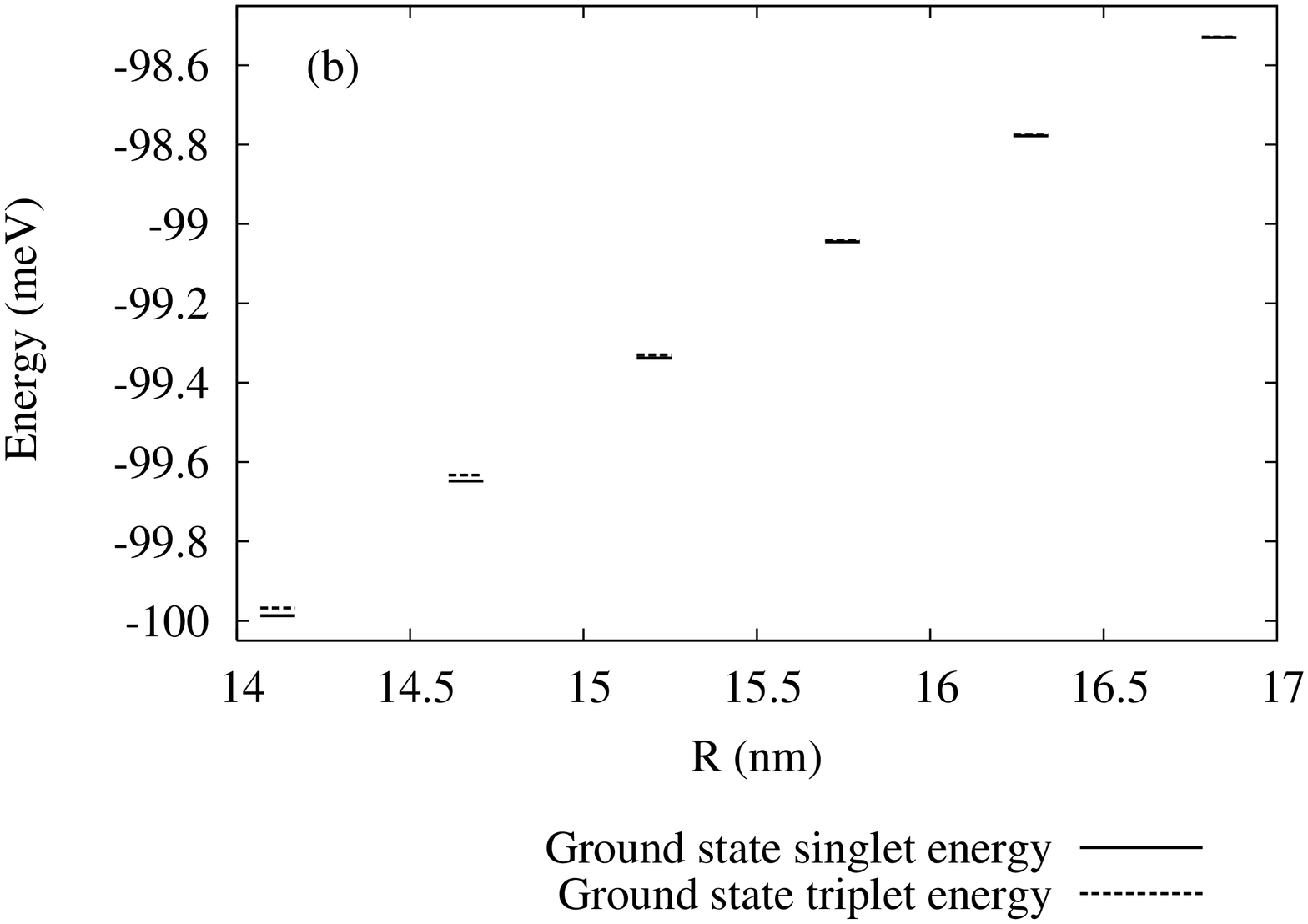} 
 \includegraphics[width=2in,angle=270]{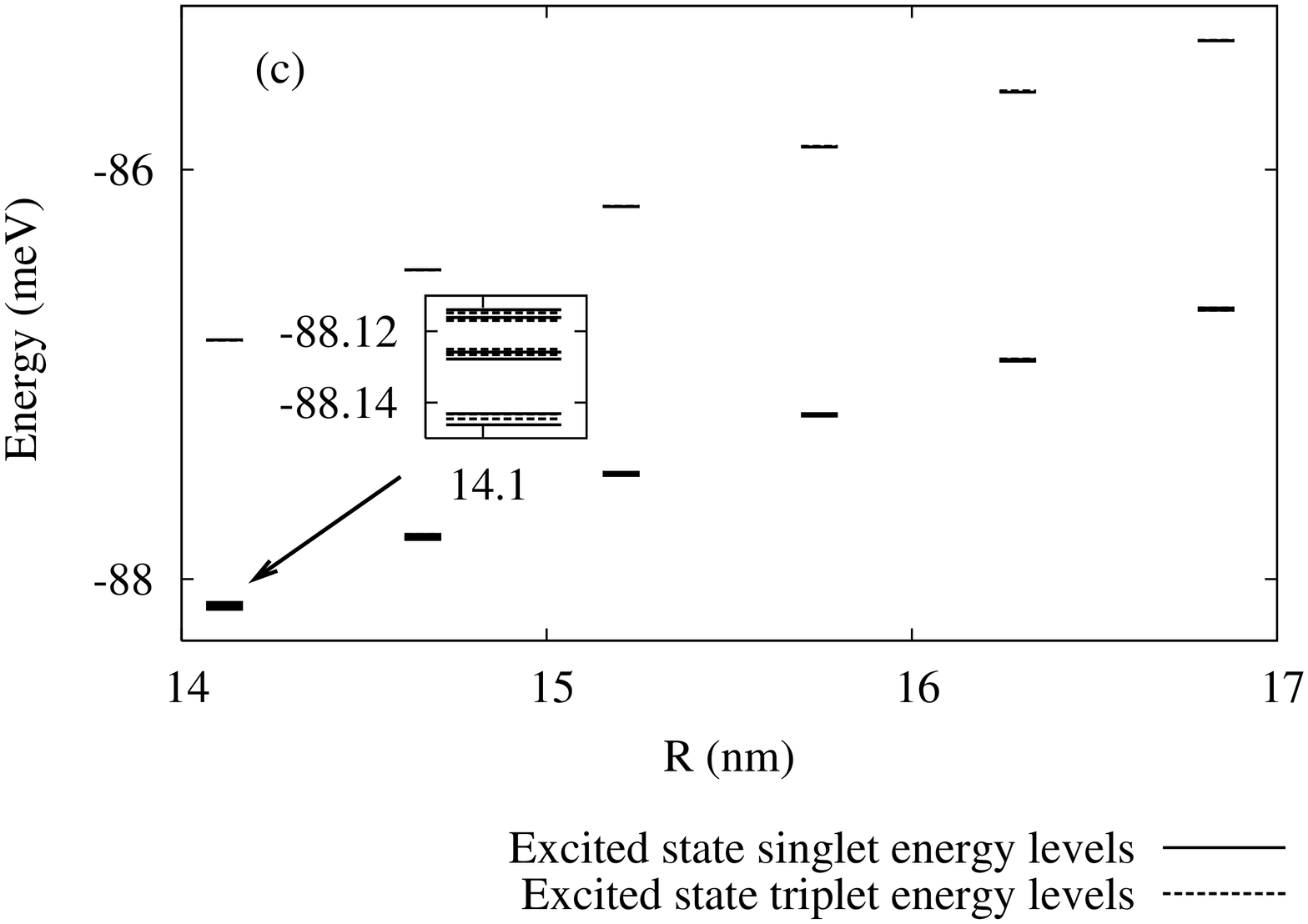}
\end{center}
 \caption{\label{fig:fign1} We compare the exchange coupling at lattice sites along the $[010]$ or $y$ direction, calculated using our three quantum chemical models in (a) for $R > 14$nm: using the H-L states, H-M states and our extended molecular orbital basis, for zero strain. In (b) we plot the ground state singlet and triplet energy separately for clarity, and in (c) we plot the rest of the excited two-electron energy levels using our extended basis. Here we only consider values of $\bm{R}$ such that both P donors are on substitutional donor sites.}
\end{figure}

\begin{figure} [t!]
\begin{center}
 \includegraphics[width=2in,angle=270]{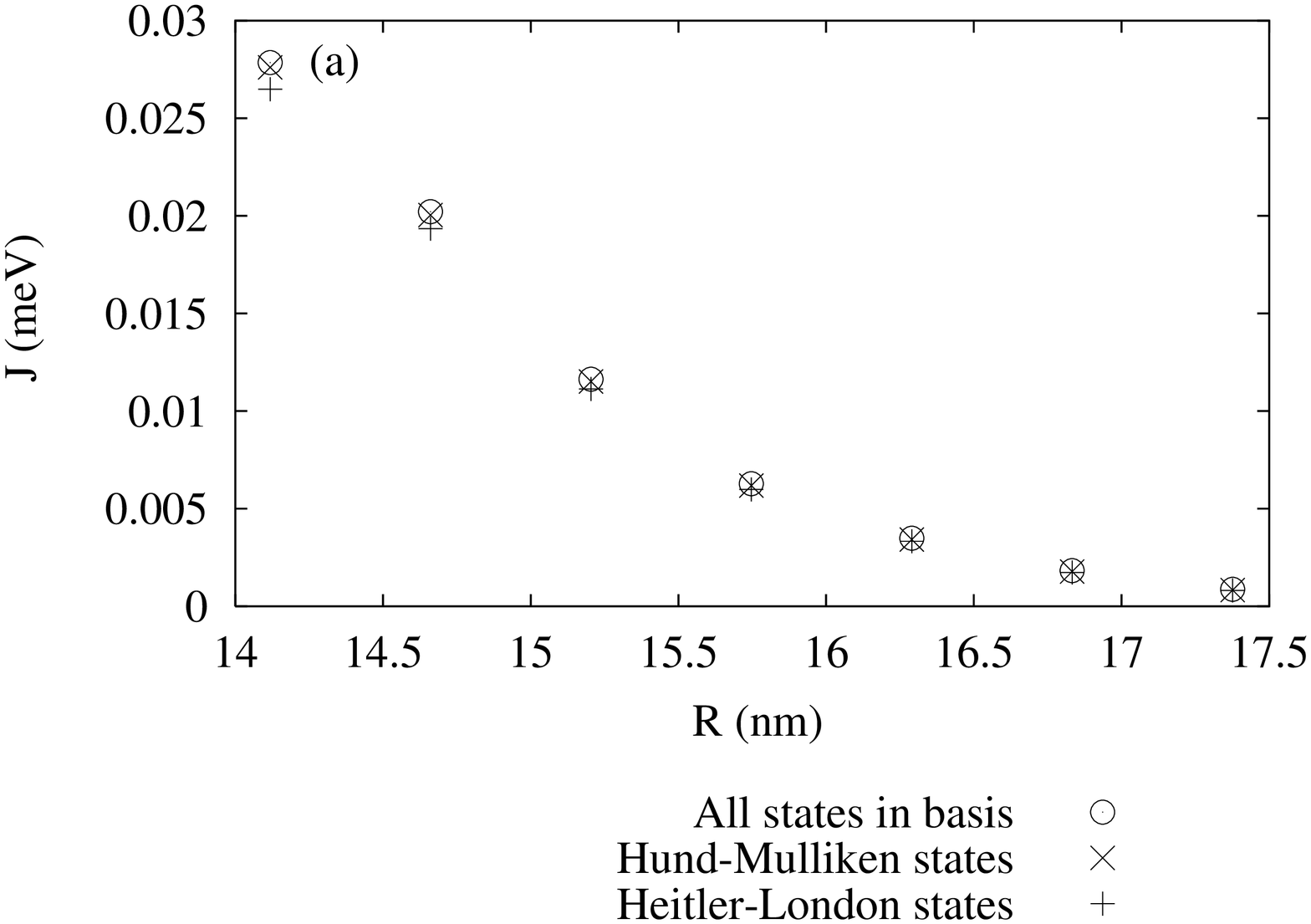} 
 \includegraphics[width=2in,angle=270]{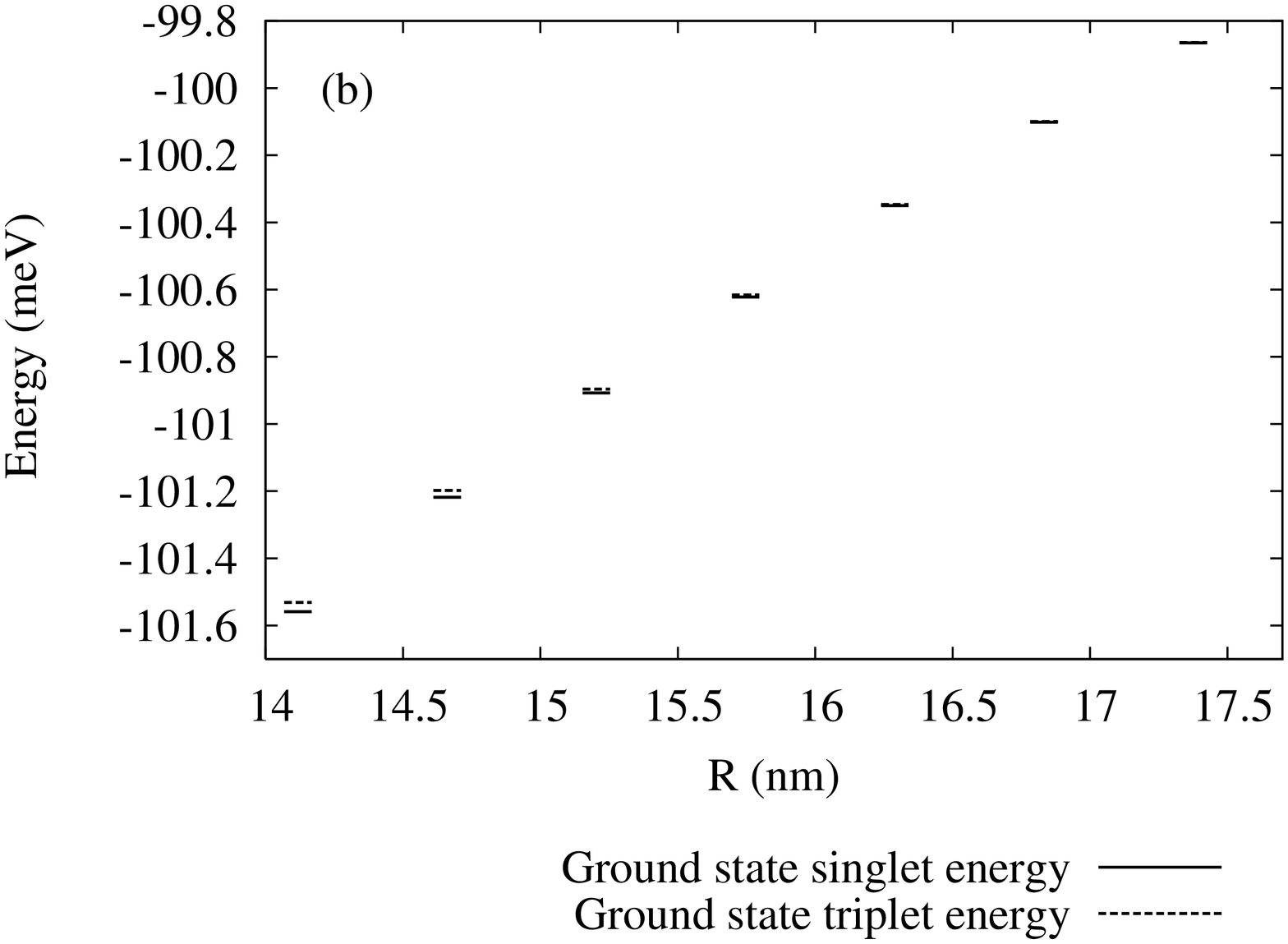} 
 \includegraphics[width=2in,angle=270]{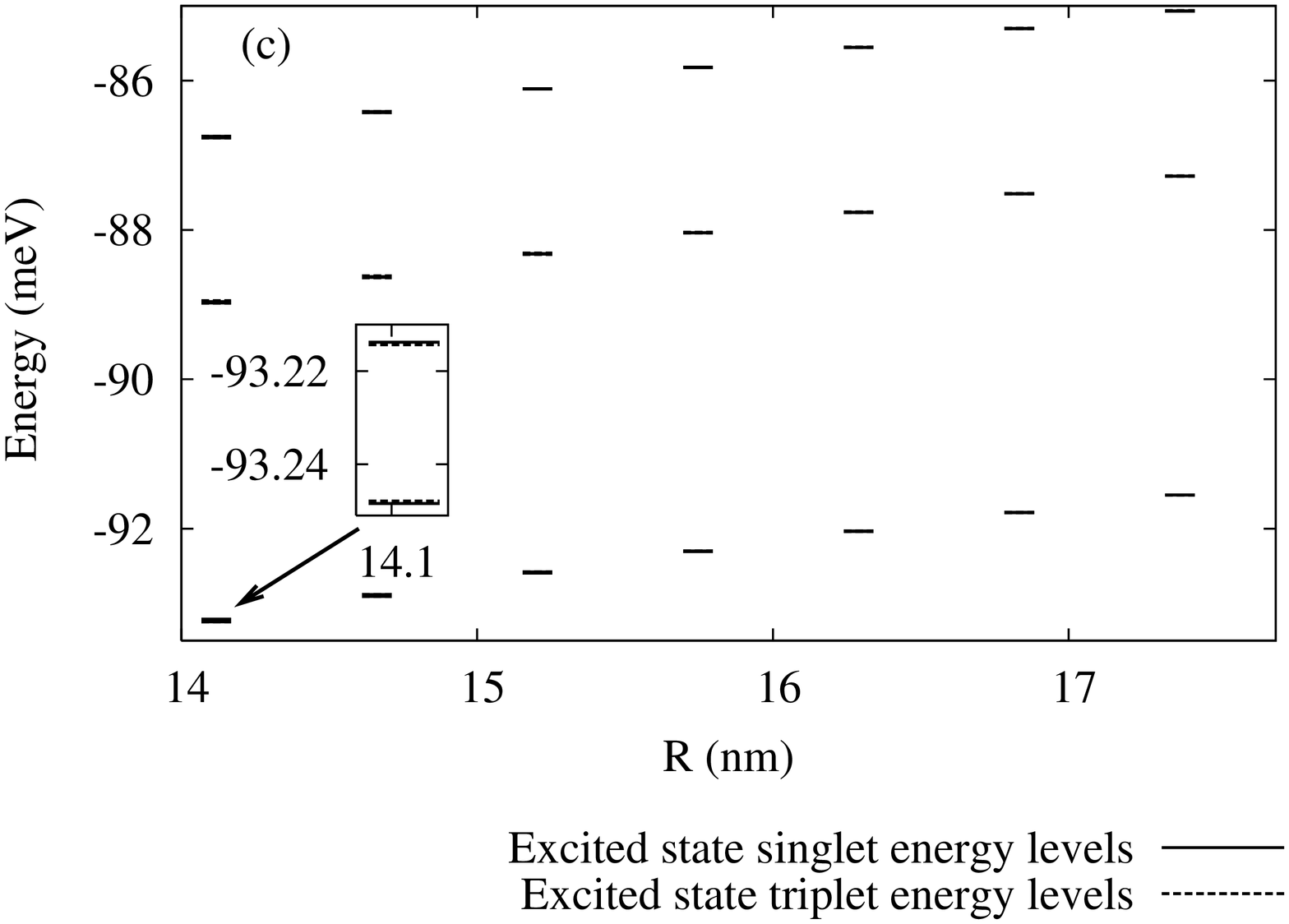}
\end{center}
 \caption{\label{fig:fign2} Comparison of the exchange coupling for $\chi=-1$ in (a) along the $[010]$ or $y$ direction. In (b) we plot the ground state and triplet state energies separately for clarity, and in (c) we plot the rest of the excited two-electron energy levels using our extended basis.  Here we only consider values of $\bm{R}$ such that both P donors are on substitutional donor sites.}
\end{figure}

\begin{figure} [h!]
\begin{center}
 \includegraphics[width=2in,angle=270]{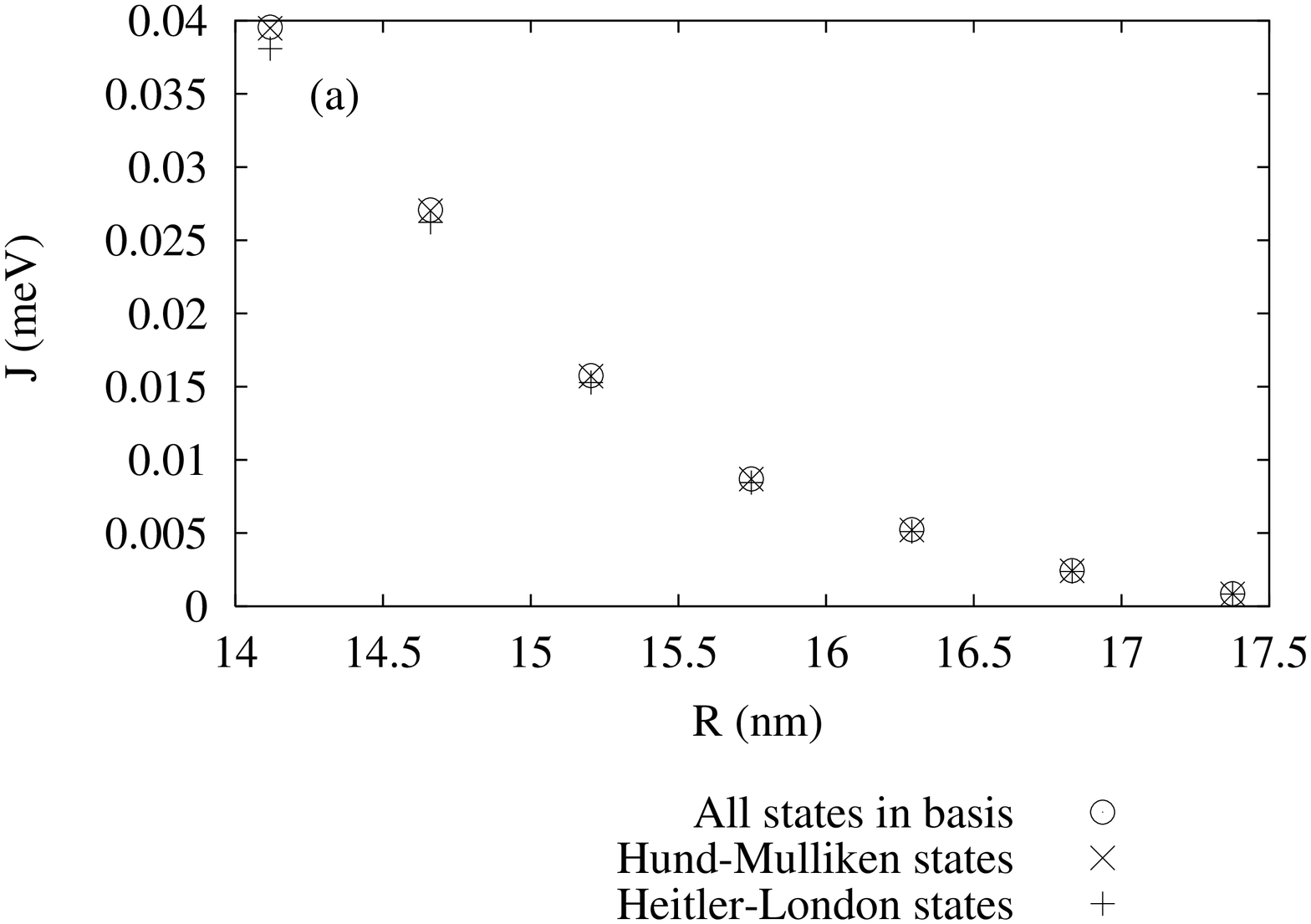} 
 \includegraphics[width=2in,angle=270]{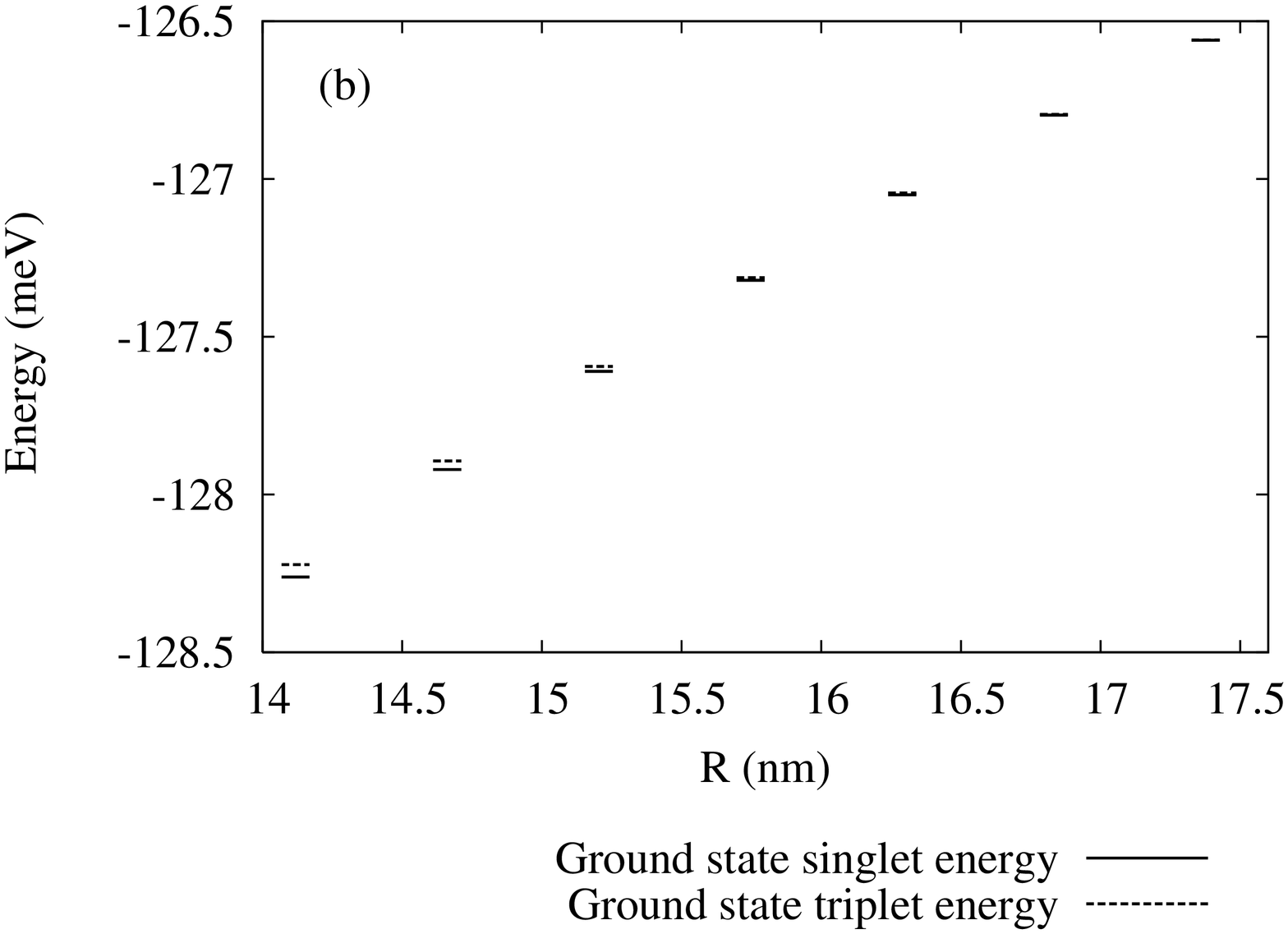} 
 \includegraphics[width=2in,angle=270]{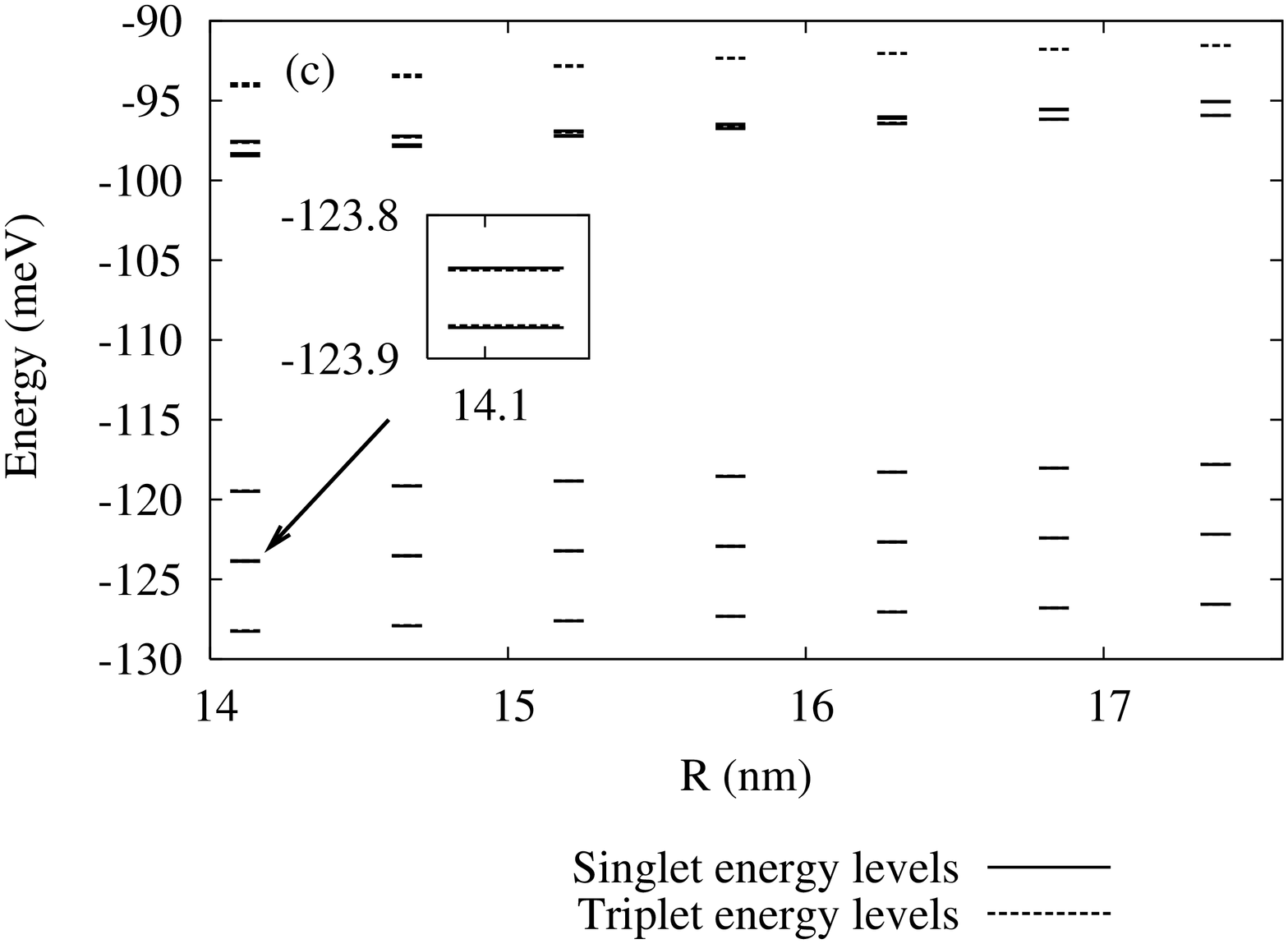}
\end{center}
 \caption{\label{fig:fign3} Comparison of the exchange coupling for $\chi=-5$ in (a) along the $[010]$ or $y$ direction. In (b) we plot the ground state and triplet state energies separately for clarity, and in (c) we plot all the singlet and triplet two-electron energy levels using our extended basis.  Here we only consider values of $\bm{R}$ such that both P donors are on substitutional donor sites.}
\end{figure}
\begin{figure} [h!]
\begin{center}
 \includegraphics[width=2in,angle=270]{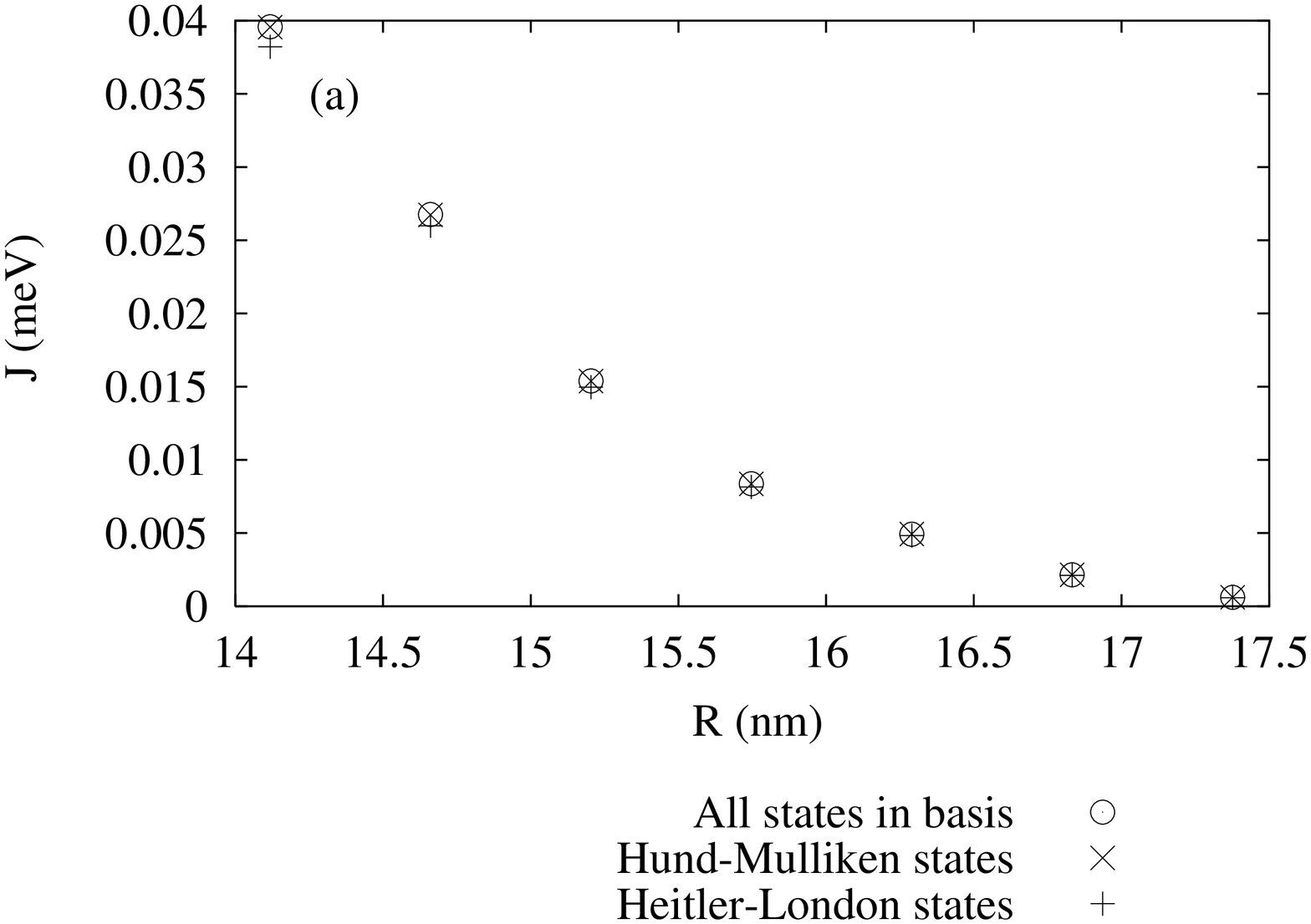} 
 \includegraphics[width=2in,angle=270]{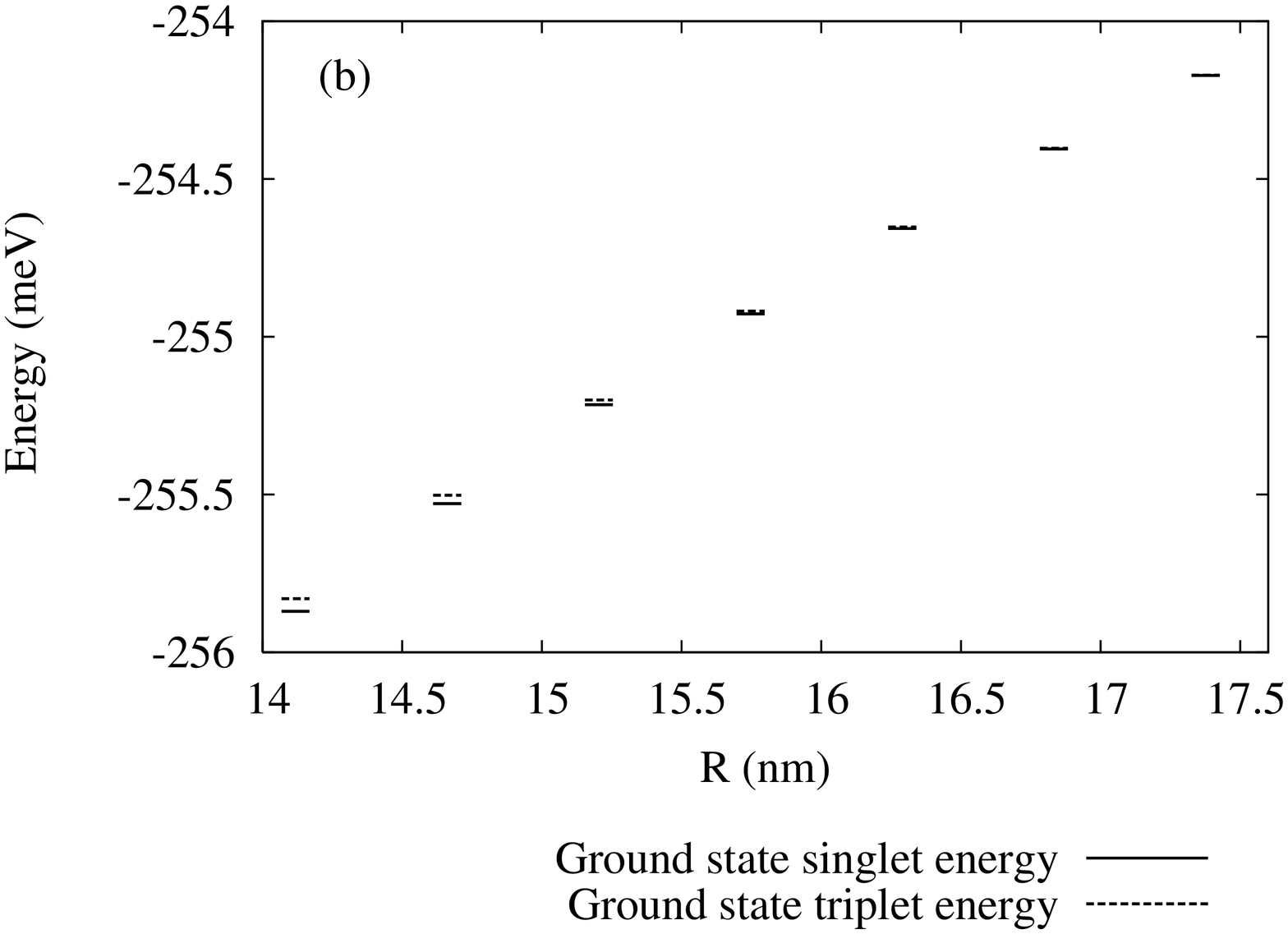} 
 \includegraphics[width=2in,angle=270]{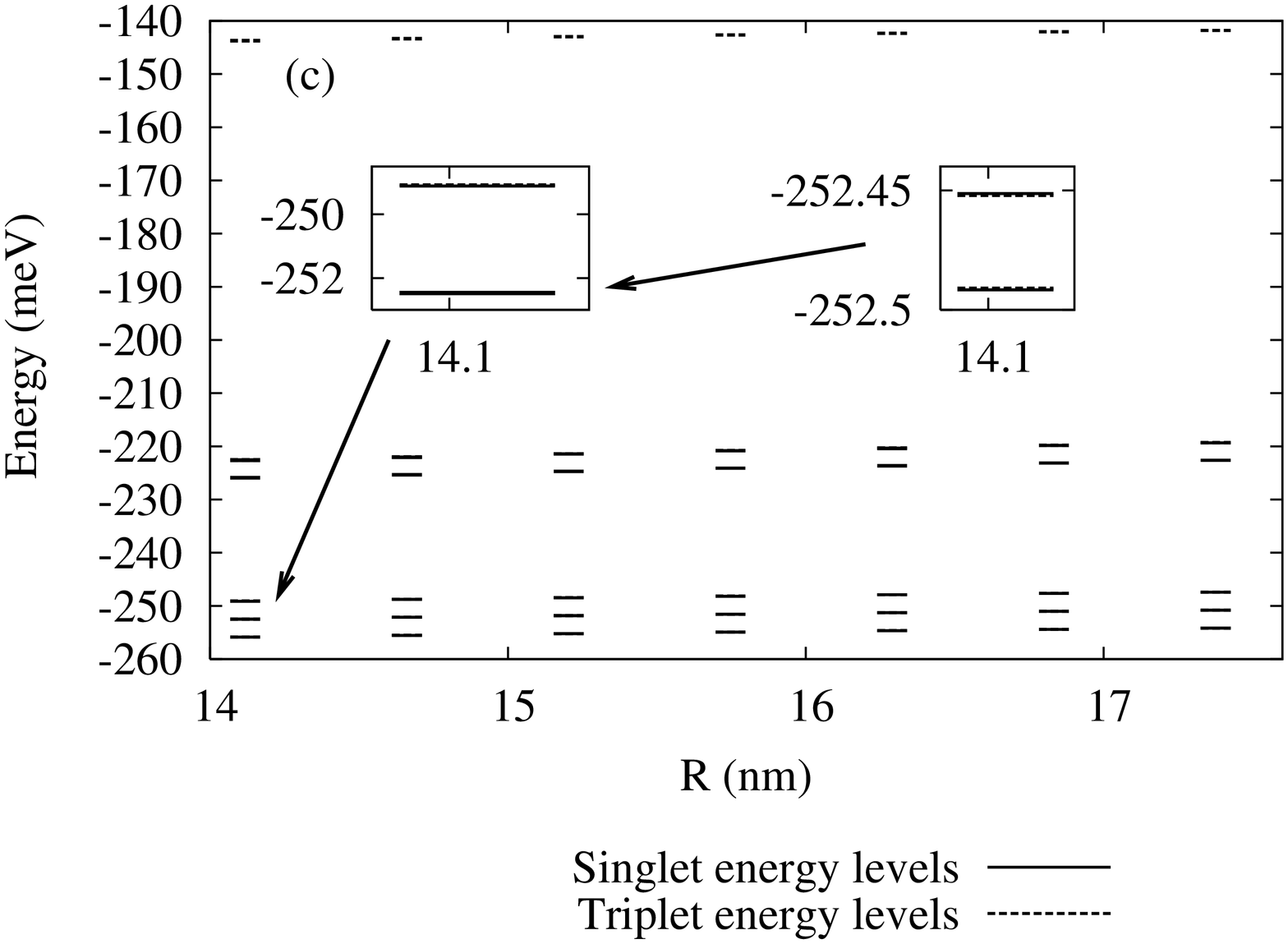}
\end{center}
 \caption{\label{fig:fign4} Comparison of the exchange coupling for $\chi=-20$ in (a) along the $[010]$ or $y$ direction. In (b) we plot the ground state and triplet state energies separately for clarity, and in (c) we plot all the singlet and triplet two-electron energy levels using our extended basis.  Here we only consider values of $\bm{R}$ such that both P donors are on substitutional donor sites.}
\end{figure}

We give the results for our three quantum chemical models for the two-electron states: the two H-L states, the four H-M states and our extended molecular orbital basis. Figures~\ref{fig:fign1a}(a) to \ref{fig:fign4}(a) show a comparison of the exchange coupling obtained using our three methods, for varying inter donor separations, and for $Q_1$ and $Q_2$ located at lattice sites. The reason why we study the exchange coupling in more detail for $R$ greater than 14nm in Fig.~\ref{fig:fign1}(a) to \ref{fig:fign4}(a) is because separations of about at least 14nm are envisioned to be needed in order for metallic gates to be placed on top of and between adjacent qubits (currently the smallest width of the metallic gates that can be fabricated is about 10nm).\cite{kane} These gates provide additional tuning of the electron density and P nuclear spin via the application of varying voltages to them. Here we consider the inter donor separations along the $y$ or $[010]$ direction only (see Fig.~\ref{fig:fig1}). This is because executing the full molecular orbital calculation is very computationally expensive. In the next section we use the H-M method to calculate the exchange coupling for many different orientations of $Q_1$ and $Q_2$ in the lattice.

We can observe that as $R$ increases the H-L calculation is more accurate as the two donors become further separated, and it becomes a better approximation to treat the two donors as a superposition of the single electron ground state wave functions centered at each donor. We demonstrate this in Fig.~\ref{fig:fign1a}(a) and \ref{fig:fign1}(a)  using smaller inter donor separations ($R \leq 12$nm), and larger inter donor separations ($R > 12$nm) respectively, for $\chi = 0$. We found that the exchange coupling is improved substantially for the smaller inter donor separations using our full molecular orbital calculations. We can also see that even if we just include the ``doubly occupied'' states in our H-M calculation we get a significant improvement in the exchange energy over H-L theory, when we compare it with the full molecular orbital calculation.

In part (b) of Fig.~\ref{fig:fign1} to \ref{fig:fign4} we show the exchange splitting between the ground singlet and triplet states using the full molecular orbital calculations. The results for the two P donor-pair wavefunction give a two-electron ground state of the order -98 to -100meV for $R > 14$nm and $\chi = 0$, (shown in Fig.~\ref{fig:fign1}(b)). The energy of two isolated P atoms should be in the order of -91meV. When we calculate the singlet and triplet energy levels, the 6-D and 3-D integrals involve both repulsive, direct Coulomb and attractive exchange integrals. We see that the attractive terms between the exchange charges and the nuclei outweigh the repulsive terms, and we obtain a molecular binding energy that is deeper than the sum of the energy of the two isolated P atoms.\cite{slater}  The single donor $A_1$ symmetry state wave functions that form the``Heitler-London" two-electron ground state, are not spherically symmetric, and the electron density for these orbitals are more heavily weighted along the co-ordinate axes (see Fig.~\ref{fig:s1}).  As a result, even at large interdonor separations for $14\mbox{nm} < R < 18$nm (along the $y$ direction), the ground state energy of the two P donor-pair wavefunction is lowered by 7 to 9meV, from that of the isolated P atoms.

Furthermore, in part (c) of these figures, and Fig.~\ref{fig:fign1a}(b) we show the energy level spectrum we calculate for our two-electron system, using our extended basis for the singlet and triplet states. In these plots the difference between the first set of excited energy levels cannot be resolved, so we have included an inset which magnifies this region. For clarity, we only plot the first eight energy levels for both the singlet and triplet two-electron bases. An interesting feature of these singlet and triplet energy levels is that the four cases of strain give very different spectra for the higher energy levels. These plots demonstrate that the ground singlet and triplet states are well separated from the rest of the higher excited states in Hilbert space, for all values of the strain parameter, $\chi$. This is because the ground singlet and triplet states are formed from the symmetric and anti-symmetric combinations of the single donor ground ``$A_1$'' states, which are much lower in energy than the next excited single donor states, the triplet $T_2$ and doublet $E$ states. However as $\chi$ decreases this energy gap becomes smaller as the single donor ground state is no longer a symmetric combination of the six conduction band states.

Table~\ref{table1} shows the difference between the ground triplet state and the first excited singlet state. Here the first excited singlet state corresponds to two-electron orbitals formed using symmetric combinations of the donor electrons at both P donors, and there is negligible contribution from the doubly occupied orbitals. These results are in full accordance with the single donor results reported earlier in Table~\ref{table2}. We can see clearly the trend in the plots, that as the strain parameter decreases the energy gap becomes smaller. However these energy gaps all remain much larger than the exchange coupling between the ground singlet and triplet states, and much higher than $k_B T \approx 0.1$meV, at the cryogenic temperatures required for quantum computing. Thus we can consider that our targeted Hilbert space, the H-L states, are well separated from the rest of the excited Hilbert space.

\begin{table}[h!]\caption{\label{table1} Energy gap between the ground triplet state and the first excited singlet state, $\Delta E$ (meV).}
\vspace{0.1cm}
\begin{center}
\begin{tabular}{|c|c|c|} \hline
$\chi$ & $R=14.118$nm  & $R = 17.376$nm \\ \hline
0 & 11.822 & 11.808 \\
-1 & 8.283 & 8.315 \\
-5 & 4.343 & 4.380 \\
-20 & 3.339 & 3.378 \\ \hline
\end{tabular}
\end{center}
\end{table}

\begin{table}[h!]\caption{\label{tableref1b} Singlet and triplet energy levels and corresponding two-electron eigenstates for $R=7.602$nm and $\chi = 0$.}
\vspace{0.1cm}
\begin{center}
\begin{tabular}{|c|c|c|} \hline
\multicolumn{3}{|c|}{$\chi = 0$ , $R=7.602$nm}  \\ \hline
\multicolumn{3}{|c|}{Singlet energy levels}  \\ \hline
 Energy  & Two-electron  & One-electron \\
(meV)  & basis states  & states involved \\ \hline
-109.18685 & $\Psi^S_{43}$  &  $A_1$ \\ 
 \multicolumn{2}{|c|}{Doubly occupied  $\Psi^S_{1}$/$\Psi^S_{22}$} &   \\ \hline
-97.77240 & $\Psi^S_{44}$/$\Psi^S_{49}$/$\Psi^S_{46}$/$\Psi^S_{61}$ & $A_1$/$T_2$($x$)  \\
 \multicolumn{2}{|c|}{Doubly occupied  $\Psi^S_{2}$/$\Psi^S_{4}$/$\Psi^S_{23}$/$\Psi^S_{25}$} &  and $A_1$/$T_2$($z$) \\ \hline
-97.76809 & $\Psi^S_{44}$/$\Psi^S_{49}$/$\Psi^S_{46}$/$\Psi^S_{61}$ & $A_1$/$T_2$($x$)  \\
 \multicolumn{2}{|c|}{Doubly occupied  $\Psi^S_{2}$/$\Psi^S_{4}$/$\Psi^S_{23}$/$\Psi^S_{25}$} &  and $A_1$/$T_2$($z$) \\ \hline
-96.38458 & $\Psi^S_{47}$/$\Psi^S_{48}$/$\Psi^S_{67}$/$\Psi^S_{73}$ & $A_1$/$E$ \\
 \multicolumn{2}{|c|}{Doubly occupied  $\Psi^S_{5}$/$\Psi^S_{6}$/$\Psi^S_{26}$/$\Psi^S_{27}$} &   \\ \hline
-96.32310 & $\Psi^S_{45}$/$\Psi^S_{55}$/$\Psi^S_{67}$/$\Psi^S_{73}$/$\Psi^S_{48}$ & $A_1$, $A_1$/$E$ \\
 \multicolumn{2}{|c|}{Doubly occupied  $\Psi^S_{3}$/$\Psi^S_{24}$/$\Psi^S_{22}$/$\Psi^S_{1}$} & and $A_1$/$T_2$($y$)  \\ \hline
-95.94663 & \multicolumn{2}{|l|}{$\Psi^S_{43}$/$\Psi^S_{47}$/$\Psi^S_{48}$/$\Psi^S_{67}$/$\Psi^S_{73}$/$\Psi^S_{55}$ }  \\                                                                  
\multicolumn{2}{|c|}{Doubly occupied } & $A_1$, $A_1$/$E$  \\
\multicolumn{2}{|c|}{$\Psi^S_{5}$/$\Psi^S_{6}$/$\Psi^S_{26}$/$\Psi^S_{27}$/$\Psi^S_{22}$/$\Psi^S_{1}$} &  and $A_1$/$T_2$($y$) \\ \hline
\multicolumn{3}{|c|}{Triplet energy levels}  \\ \hline
 Energy  & Two-electron  & One-electron \\
(meV)  & basis states  & states involved \\ \hline
-106.90440 & $\Psi^T_{31}$  &  $A_1$ \\ \hline
-97.77305 & $\Psi^T_{32}$/$\Psi^T_{37}$  & $A_1$/$T_2$($z$)  \\
 \multicolumn{2}{|c|}{Doubly occupied  $\Psi^T_{1}$/$\Psi^T_{16}$} &   \\ \hline
-97.76847 & $\Psi^T_{34}$/$\Psi^T_{49}$  & $A_1$/$T_2$($x$)  \\
 \multicolumn{2}{|c|}{Doubly occupied  $\Psi^T_{3}$/$\Psi^T_{18}$} &   \\ \hline
-96.50967 & $\Psi^T_{35}$/$\Psi^S_{36}$/$\Psi^T_{55}$/$\Psi^T_{61}$  & $A_1$/$E$  \\
 \multicolumn{2}{|c|}{Doubly occupied  $\Psi^T_{4}$/$\Psi^T_{5}$/$\Psi^T_{19}$/$\Psi^T_{20}$} &   \\ \hline
-96.19400 & $\Psi^T_{33}$/$\Psi^T_{43}$  & $A_1$/$T_2$($y$)  \\
 \multicolumn{2}{|c|}{Doubly occupied  $\Psi^T_{2}$/$\Psi^T_{17}$} &   \\ \hline
-95.76107 & $\Psi^T_{33}$/$\Psi^S_{43}$/$\Psi^T_{55}$  & $A_1$/$T_2$($y$)  \\
 \multicolumn{2}{|c|}{Doubly occupied  $\Psi^T_{2}$/$\Psi^T_{5}$/$\Psi^T_{17}$} & and $A_1$/$E$   \\ \hline
\end{tabular}
\end{center}
\end{table}

\begin{table}[h!]\caption{\label{tableref1} Singlet and triplet energy levels and corresponding two-electron eigenstates for $R=14.118$nm and $\chi = 0$.}
\vspace{0.1cm}
\begin{center}
\begin{tabular}{|c|c|c|} \hline
\multicolumn{3}{|c|}{$\chi = 0$ , $R=14.118$nm}  \\ \hline
\multicolumn{3}{|c|}{Singlet energy levels}  \\ \hline 
 Energy  & Two-electron  & One-electron \\ 
(meV)  & basis states  & states involved \\ \hline 
-99.98734 & $\Psi^S_{43}$  & $A_1$ \\ \hline
-88.14626 & $\Psi^S_{46}$/$\Psi^S_{61}$ & $A_1$/$T_2$($x$) \\
-88.14312 & $\Psi^S_{44}$/$\Psi^S_{49}$ & $A_1$/$T_2$($z$) \\ \hline
-88.12780 & $\Psi^S_{45}$/$\Psi^S_{55}$ & $A_1$/$T_2$($y$) \\
-88.12567 & $\Psi^S_{45}$/$\Psi^S_{55}$ &  \\ \hline 
-88.11610 & $\Psi^S_{44}$/$\Psi^S_{46}$/$\Psi^S_{49}$/$\Psi^S_{61}$ & $A_1$/$T_2(z)$  \\
-88.11399 & $\Psi^S_{44}$/$\Psi^S_{46}$/$\Psi^S_{49}$/$\Psi^S_{61}$ & and $A_1$/$T_2$ ($x$)\\ \hline 
-86.83179 & $\Psi^S_{48}$/$\Psi^S_{67}$/$\Psi^S_{73}$ & $A_1$/$E$ \\
-86.82667 & $\Psi^S_{47}$/$\Psi^S_{48}$/$\Psi^S_{67}$/$\Psi^S_{73}$ & \\ 
-86.82358 & $\Psi^S_{47}$/$\Psi^S_{48}$/$\Psi^S_{67}$ &  \\ 
-86.80250 & $\Psi^S_{47}$/$\Psi^S_{48}$/$\Psi^S_{67}$/$\Psi^S_{73}$ & \\ \hline \hline
\multicolumn{3}{|c|}{Triplet energy levels}  \\ \hline
 Energy  & Two-electron  & One-electron \\
(meV)  & basis states  & states involved \\ \hline
-99.96806 & $\Psi^T_{31}$  & $A_1$ \\ \hline
-88.14459 & $\Psi^T_{34}$/$\Psi^T_{49}$ & $A_1$/$T_2$($x$) \\
-88.14316 & $\Psi^T_{32}$/$\Psi^T_{37}$ & $A_1$/$T_2$($z$) \\ \hline
-88.12656 & $\Psi^T_{33}$/$\Psi^T_{43}$ & $A_1$ /$T_2$($y$) \\
-88.12504 & $\Psi^T_{33}$/$\Psi^T_{43}$ &  \\ \hline
-88.11693 & $\Psi^T_{34}$/$\Psi^T_{49}$ & $A_1$/$T_2$ ($x$) \\
-88.11485 & $\Psi^T_{32}$/$\Psi^T_{37}$ & $A_1$/$T_2$($z$)  \\ \hline
-86.83227 & $\Psi^T_{35}$/$\Psi^T_{36}$/$\Psi^T_{55}$/$\Psi^T_{61}$ & $A_1$/$E$  \\
-86.81847 & $\Psi^T_{35}$/$\Psi^T_{36}$/$\Psi^T_{55}$/$\Psi^T_{61}$ & \\
-86.79819 & $\Psi^T_{35}$/$\Psi^T_{36}$/$\Psi^T_{55}$/$\Psi^T_{61}$ &   \\
-86.79310 & $\Psi^T_{35}$/$\Psi^T_{36}$/$\Psi^T_{55}$/$\Psi^T_{61}$  & \\ \hline
\end{tabular}
\end{center}
\end{table}

\begin{table}[h!]\caption{\label{tableref1a} Singlet energy levels and corresponding two-electron eigenstates for $R=14.118$nm and $\chi = -20$.}
\vspace{0.1cm}                                                                                            \begin{center}                                                                                            \begin{tabular}{|c|c|c|} \hline
\multicolumn{3}{|c|}{$\chi = -20$ , $R=14.118$nm}  \\ \hline
 Energy  & Two-electron  & One-electron \\
(meV)  & basis states  & states involved \\ \hline
-255.87009 & $\Psi^S_{43}$  & $\approx \left( F_1^{(z)}(1S) + F_1^{(-z)}(1S) \right)/\sqrt{2}$ \\ \hline
-252.49159 & $\Psi^S_{44}$/$\Psi^S_{49}$ & $\approx F_1^{(-z)}(1S)  $ \\
-252.45132 & $\Psi^S_{44}$/$\Psi^S_{49}$ & $ \approx  F_1^{(z)}(1S) $ \\ \hline
-249.11343 & $\Psi^S_{50}$ & $\approx \left( F_1^{(z)}(1S) - F_1^{(-z)}(1S) \right)/\sqrt{2}$ \\ \hline
-225.97320 & $\Psi^S_{1}$/$\Psi^S_{22}$ & Doubly occupied   \\ 
-225.82787 & $\Psi^S_{1}$/$\Psi^S_{22}$ &  $ \left( F_1^{(z)}(1S) + F_1^{(-z)}(1S) \right) /2 $  \\ \hline
\end{tabular}
\end{center}
\end{table}

Tables \ref{tableref1b} and \ref{tableref1} list the lowest set of singlet and triplet two-electron eigenvalues and eigenvectors for $\chi = 0$ and $R=7.602$ and $14.118$nm respectively, and the corresponding single donor basis states contributing. Similarly, table \ref{tableref1a} shows the singlet two-electron states for $\chi = -20$. These tables clearly show the difference in the eigenvector basis components with and without strain applied. For $R=14.118$nm, the ground singlet state is the Heitler-London state $\Psi^S_{43}$, which is composed of the single donor ground states at $Q_1$ and $Q_2$. But we see in table \ref{tableref1a} for $\chi = -20$ that this single donor ground state is no longer the six-valley degenerate $A_1$ symmetry state, as the strain has broken the degeneracy of the six valleys. The lowest energy states are when the effective Bohr radius in the direction parallel to the strain is decreased, ie. the single donor $F_1^{(\pm z)}(1S)$ basis states.

For $\chi = 0$ the first two singlet and triplet excited states are nearly degenerate and these two-electron eigenstates involve significant contributions from the single donor $A_1$ and $T_2$ symmetry states. Here we see in Table \ref{tableref1b} and \ref{tableref1} that for $\chi = 0$ these two states involve the $T_2$ states in the $\pm x$ and $\pm z$ valleys, and are lower in energy than the next states involving the $T_2$ states in the $\pm y$ valleys. This is because the overall two-electron/two-nuclei coupling leads to a more stable configuration when the donor electron densities are centered toward each other along the inter donor axis ($y$-axis), ie.~the $T_2(x)$ and $T_2(z)$ states. In contrast, Table \ref{tableref1a} shows that for $\chi = -20$ that all the higher excited states involve significant contributions from the $F_1^{(\pm z)}(1S)$ basis states. This is because with such a large strain applied, the lowest single-donor eigenstates are when the effective Bohr radius in the direction parallel to the strain is decreased, ie. the single donor $F_1^{(\pm z)}(1S)$ basis states. 

Tables \ref{tableref1b} and \ref{tableref1} compare the degeneracy lifting of the $T$ and $E$ states for small and large inter donor separations. This may be interesting for or relevant to the proposed optical Raman experiments in the unstrained case.\cite{koiller3} We find that lowest few energy excited states are formed when $A_1$ mixes with either $T_2$ or $E$ symmetry states. For $R=7.602$nm, the two-electron $A_1/E$ state is even lower in energy than the $A_1/T_2(y)$ state. As we noted earlier this is because the $T_2(y)$ state has a smaller effective Bohr radius along the inter donor axis. The two-electron donor pair is more stable when the electron wave functions are more centered toward each other along the inter donor axis. For example, at the energies corresponding to the states containing the $E$ one-electron symmetry states, the two-electron wave-functions are very mixed and we are no longer able to assign the energy levels to pure basis states.

For the smaller donor separation with $\chi=0$, and with a strain applied, the two-electron wave functions contain significant contributions from the doubly occupied orbitals. This study shows that these doubly occupied states are important basis functions to include. We have improved upon a previous study\cite{koiller3} where only H-L type orbitals were considered. Because we use an extended basis which includes both doubly occupied and H-L type two-electron orbitals, we find that the two-electron wave functions are often mixed states.


The two-electron energies given in Table \ref{tableref1a} and plotted in Fig.~\ref{fig:fign2} to \ref{fig:fign4} are not scaled so that the conduction band bottom is at zero-energy. We evaluated relative energy shifts using the valley strain parameter, $\chi$, for the single-donor strain Hamiltonian matrix and neglected any shift proportional to identity in it, to be consistent with the calculations of Koiller et al.\cite{koiller} Therefore, the calculated energies do not refer to the zero-energy to be at the bottom of the conduction band. But the energy eigenvalues shown in Table \ref{tableref1a} and Fig.~\ref{fig:fign2} to \ref{fig:fign4} give the correct relative splitting among the eigenstates.

Even just considering the four H-M states in our two-electron basis, gives an exchange energy which is very close to the extended basis calculation. For the range of device parameters we consider, the H-M calculation is the most convenient method to use since it is relatively inexpensive and very accurate. For this reason we use this method exclusively in the next section to obtain accurate results expediently and rapidly.

\subsection{Results using Hund-Mulliken basis}
\begin{figure} [t!]
\begin{center}
 \includegraphics[width=2in,angle=270]{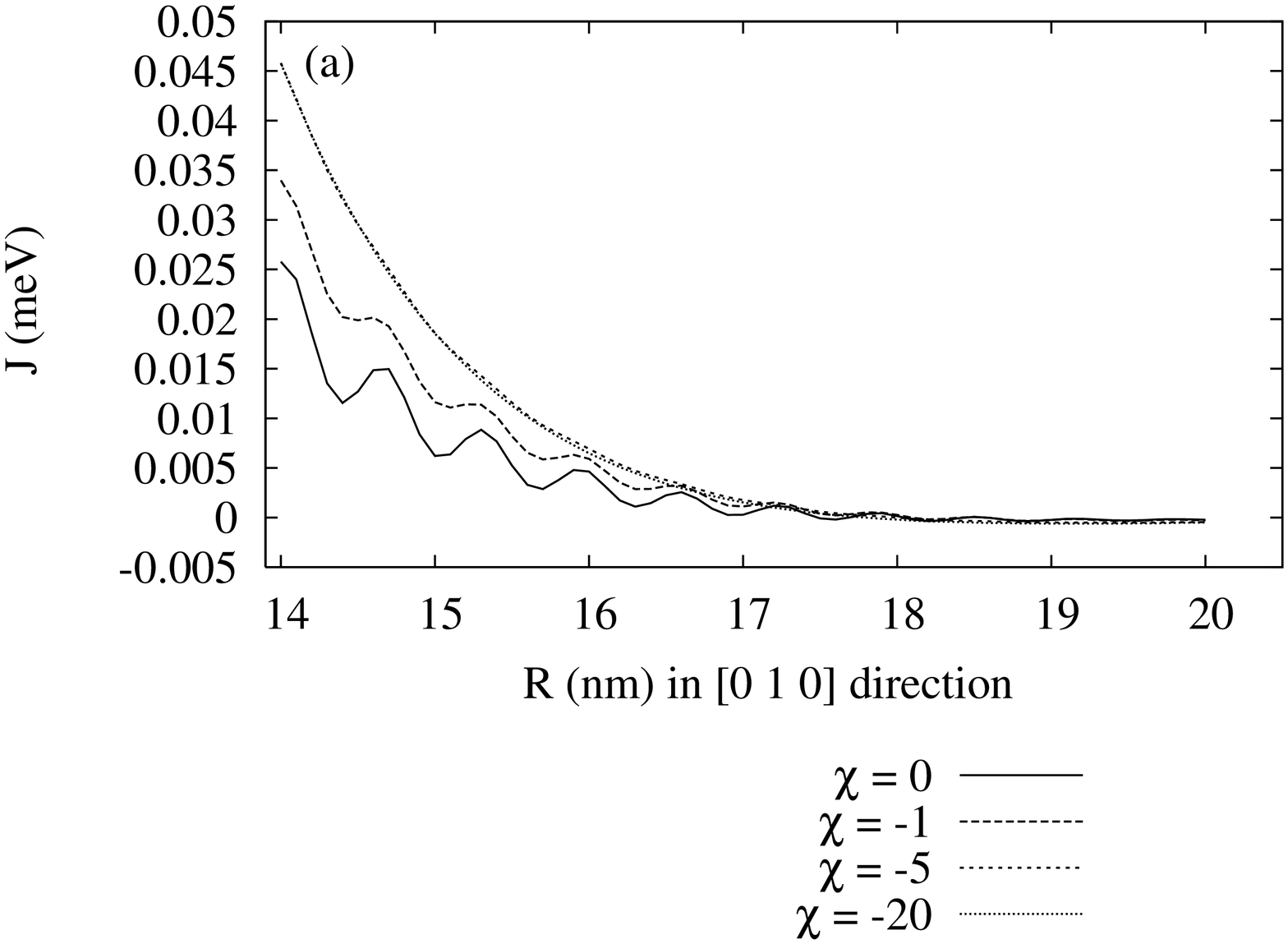} 
 \includegraphics[width=2in,angle=270]{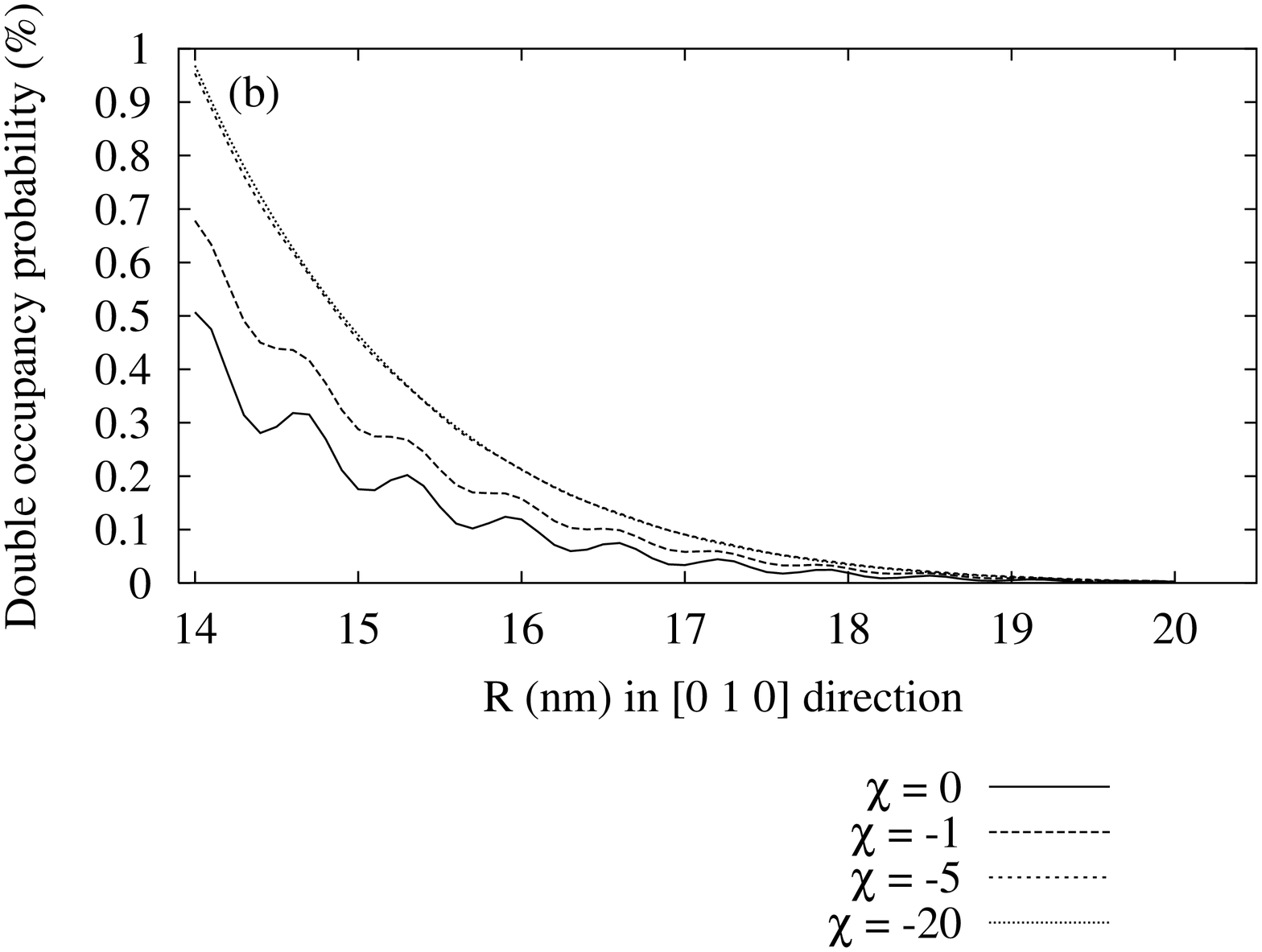} 
\end{center}
 \caption{\label{fig:figh1} Comparison of the Hund-Mulliken exchange coupling for different values of strain parameter, $\chi$, for $R$ in the $[010]$ or $y$ direction. We also plot the strain dependence of the double occupancy probability in (b).}
\end{figure}

In figure \ref{fig:figh1}(a) we show the H-M calculation for the exchange energy for a range of inter donor separations along the $[010]$ direction to compare with the previous section. This plot demonstrates the oscillations in the exchange energy due to the inter valley interference between the degenerate conduction band minima. This oscillatory nature of the exchange coupling has already been reported by Wellard \emph{et al.}\cite{cam} and Koiller \emph{et al.}\cite{koiller, koiller2} using a H-L calculation. We have improved upon and checked the H-L approximation, by extending our basis to include the H-M states, and found that the H-M basis offers some improvement for the close inter donor separations, and also allows us to calculate the ground state double occupancy probability of both electrons on the same donor.

We observe that for the zero strain case the oscillations in the exchange are the most conspicuous, as the inter valley interference is highest, because there are equal contributions from all six valleys in the ground singlet and triplet states. However, when a strain is applied in the $z$-direction, we find that the $\pm z$ valleys are favored energetically (see Fig.~\ref{fig:s2} ), which implies a larger effective Bohr radius along the inter donor axis ($y$), and hence the exchange coupling is improved substantially for this particular orientation of the P donor atoms along the $[010]$ axis.\cite{koiller} 

In addition to calculating the variation of the exchange coupling with inter donor separation in Fig.~\ref{fig:figh1}, we also calculated the probability of the ground state double occupation of the H-M singlet states ($\Psi^S_1$ and $\Psi^S_{22}$) in (b).\cite{hu} The exchange coupling with only the H-M states, matches the exchange coupling with the full spectrum of states very accurately. Thus, based on this we expect that the most appreciable contribution of the doubly occupied states will be from the H-M doubly occupied singlet states in our basis, which is why we only report the double occupancy probability of these states.

This double occupation probability is also an important parameter for quantum gate operations using electron spins as qubits. For non-zero exchange coupling the spin degrees of freedom are also correlated with the orbital motion, and there is some probability that both electrons will be on the same donor.\cite{barrett} These doubly occupied states are not part of the targeted computational space, and as such, are a potential source of leakage error in solid-state quantum computers. These plots show that as the strain parameter decreases, (and the effective Bohr radii in the $x$ and $y$-directions increases), this double occupancy probability increases as one may expect.

Schliemann \emph{et al.}\cite{schliemann} showed that gate operations on coupled quantum dot pairs which temporarily  increase the exchange splitting, in order to swap electronic spins, inevitably lead to a finite double occupancy probability for both dots. However they showed that this double occupancy amplitude does not lead to significant errors in quantum computing, provided that after the gate action is completed, the double occupancy probability is vanishingly small. But if the double occupancy probability occurs to any sizable extent before or as a result of the gating action, then any quantum computer based on this hardware is likely to fail. Thus for the fabrication of these devices we want to minimise the double occupation probability for all states at zero voltage.

The exchange coupling increases consequently with a uniaxial strain applied in a direction perpendicular to the inter donor axis, which achieves faster gating times. However, the double occupancy probability also increases correspondingly, which increases the error requirement subsequently. In the last section we saw that both these features also lead to a greater mixing of the ground state with the higher excited states, which causes the energy levels to become closer together. This energy splitting between the targeted H-L orbitals and the rest of the excited two-electron states informs us if during the gating action, the coupled donor system is well isolated and the higher excited states can be safely neglected. This can also give us an estimate for gating times, so that the gating operation remains adiabatic.

\begin{figure} [t!]
\begin{center}
 \includegraphics[width=2in,angle=270]{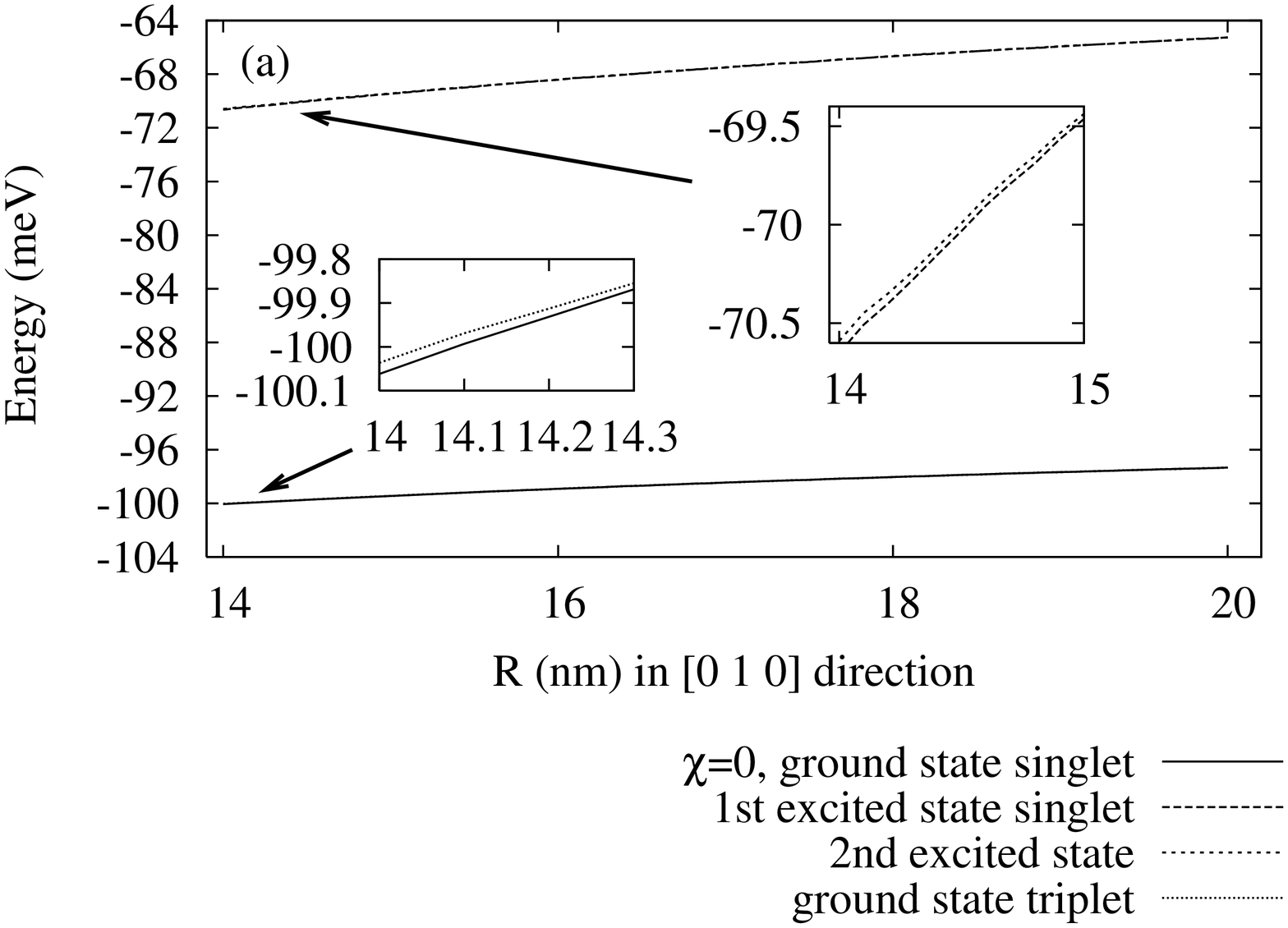} 
 \includegraphics[width=2in,angle=270]{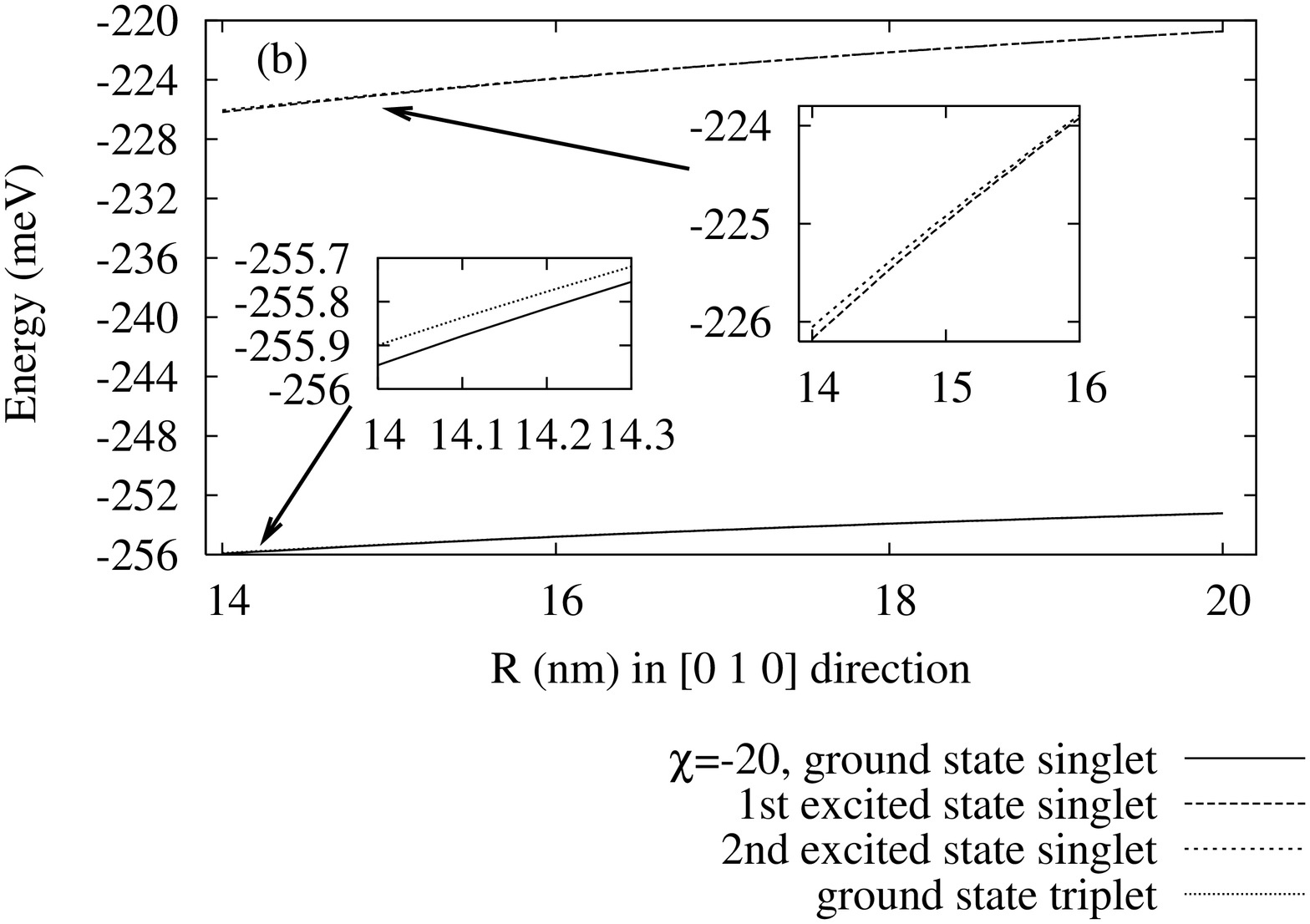} 
\end{center}
 \caption{\label{fig:figh2} Plot of the three singlet and one triplet energy levels calculated with the H-M basis, for $\chi = 0$ in (a) and $\chi=-20$ in (b).}
\end{figure}

In Fig.~\ref{fig:figh2} we plot the four two-electron energy levels predicted using our H-M basis, in order to compare with our more rigorous evaluation of the higher excited states in Fig.~\ref{fig:fign1} and \ref{fig:fign4}, parts (b) and (c). The insets in these plots magnify the splitting between the energy levels. The bottom inset in both plots corresponds to the ground state singlet and triplet energy levels, and compares favorably with the results shown for this splitting using the molecular orbital basis, in Fig.~\ref{fig:fign1}(b) and \ref{fig:fign4}(b) for $R \approx 14$nm.

We observe that although the H-M basis provides an adequate description of the ground singlet and triplet ``H-L'' states and exchange coupling, it is unable to predict the higher energy levels accurately. This is because as we reported earlier in Table \ref{tableref1}, using our full molecular orbital calculations, the first excited singlet state does not include significant contributions from the doubly occupied H-M singlet basis states. This leads to an erroneously large energy splitting between the targeted ground ``H-L'' orbitals and the higher excited states.

\begin{figure} [htp]
\begin{center}

\setlength{\unitlength}{0.15cm}
\begin{picture}(60,25)

\put(17,5){\makebox(0,0){\bfseries {$Q_1$}  }}
\put(55,5){\makebox(0,0){\bfseries {$Q_2$}  }}

\put(20.5,5){\vector(1,0){10}}

\put(20.5,5){\vector(-1,-1){7}}

\put(20.5,5){\vector(0,1) {10}}

\put(12, 0){\makebox(0,0) {$x$   }}

\put(17.5, 13){\makebox(0,0) {$z$   }}

\put(29,2){\makebox(0,0){$y$}}

\put(20.5,5){\vector(1,0) {30}}

\put(42,2){\makebox(0,0){$\mathbf{R}$}}

\qbezier(18.5,3)(22.5,2)(25.5,5)

\qbezier(20.5,8.5)(24.5,7)(25.5,5)

\put(22.5, 1){\makebox(0,0) {$\theta$   }}
\put(24.5,9){\makebox(0,0) {$\phi$   }}

\put(20.5,5){\circle{2}}
\put(50.5,5){\circle{2}}

\end{picture}
\end{center}

\caption{\label{fig:figh4} We evaluate the exchange coupling for fixed $|\bm{R}|=14$nm and small displacements of $Q_2$, about the targeted inter donor separation of $\bm{R}=(0,14\mbox{nm},0)$, by varying $\theta$ and $\phi$ in our calculations.}
\end{figure}
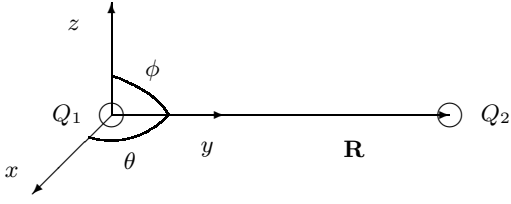

Furthermore we also investigated the variation of the exchange coupling for $Q_1$ and $Q_2$ displaced at small distances away from a targeted inter donor separation along the $[010]$ or $y$-axis. In these calculations we fixed the magnitude of the inter donor separation to be 14nm, and varied the two angular variables, $\theta$ and $\phi$ defined in Fig.~\ref{fig:figh4}.

\begin{figure} [h!]
\begin{center}
 \includegraphics[width=2in,angle=270]{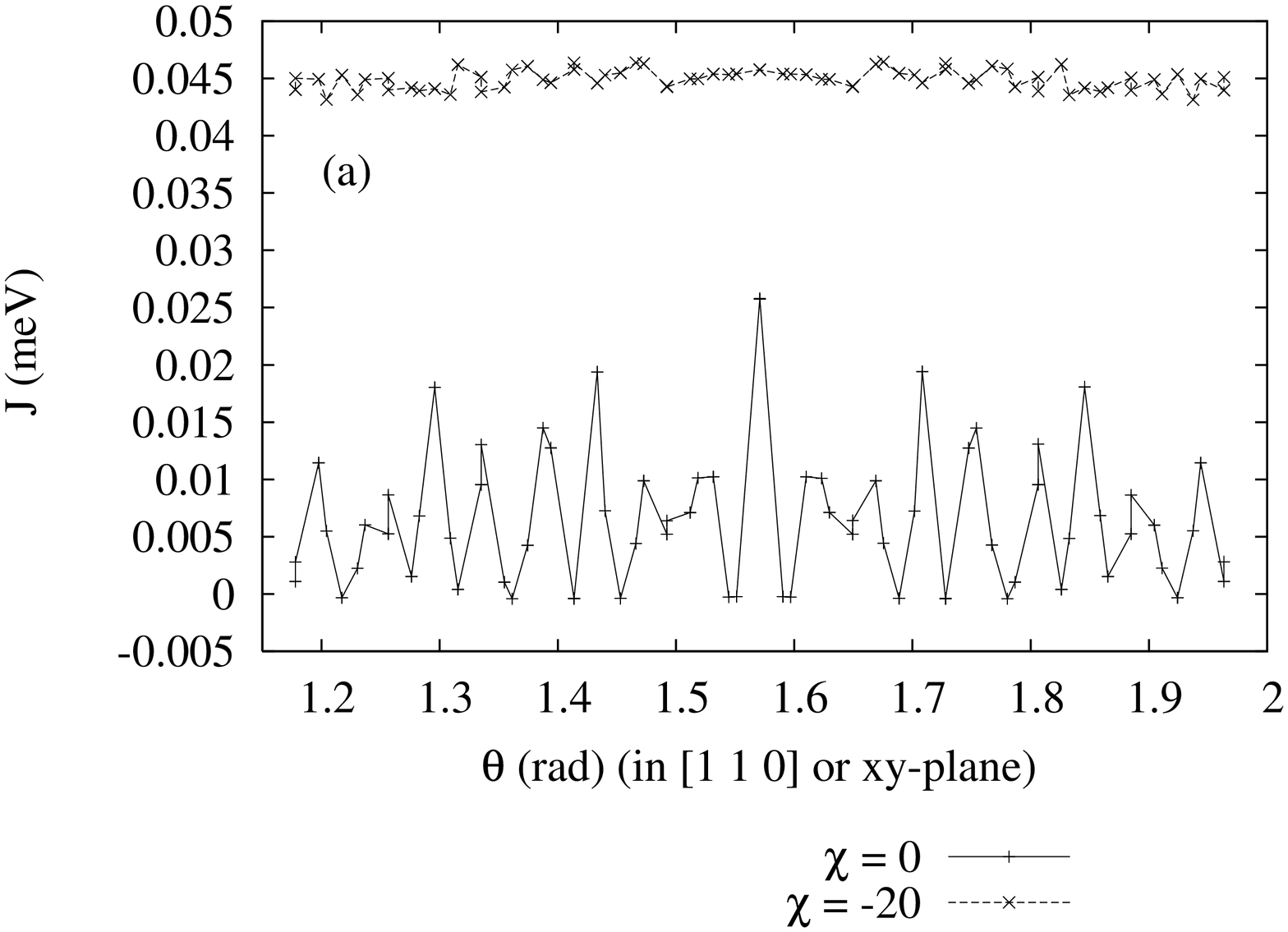} 
 \includegraphics[width=2in,angle=270]{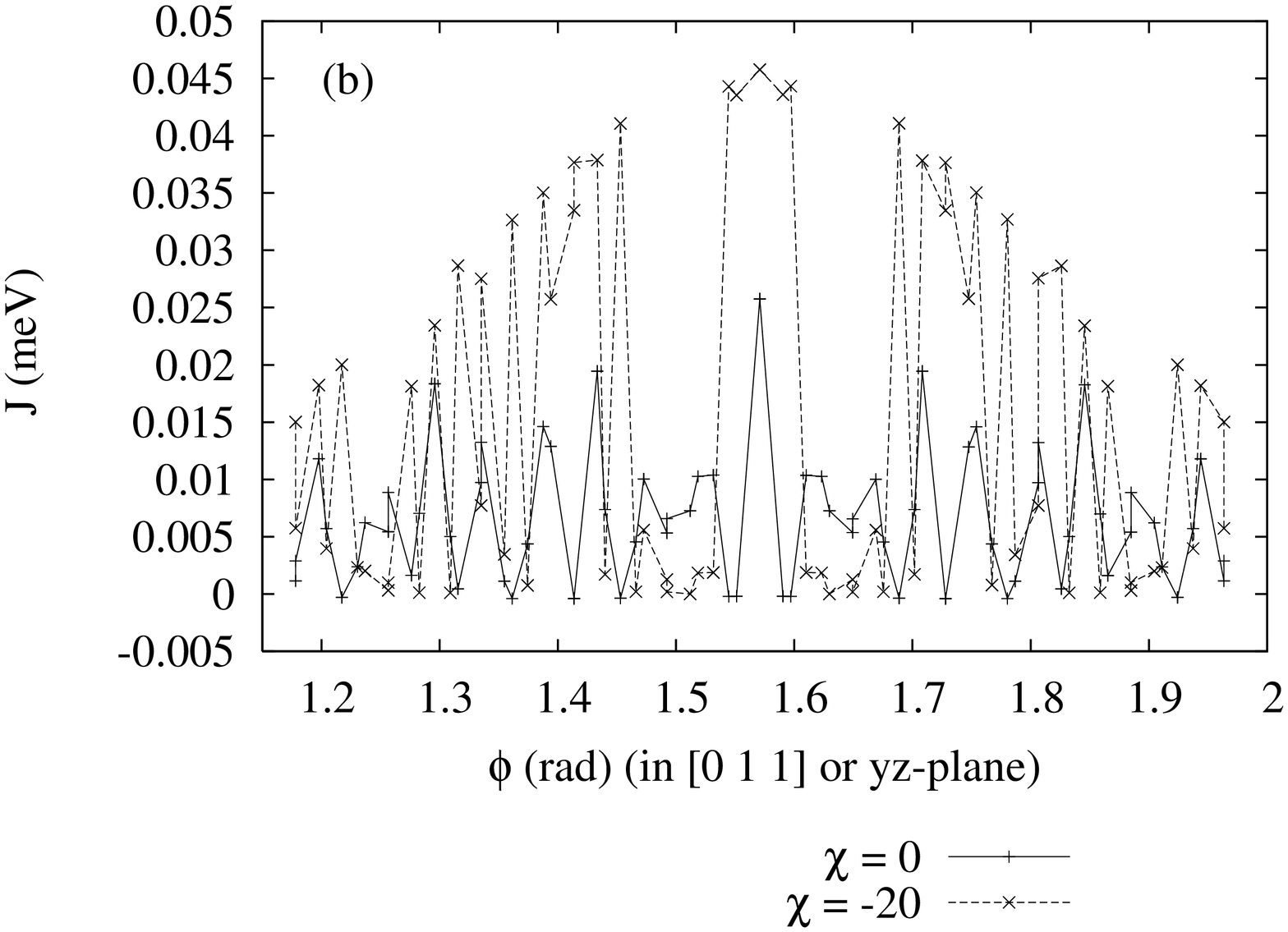} 
\includegraphics[width=2in,angle=270]{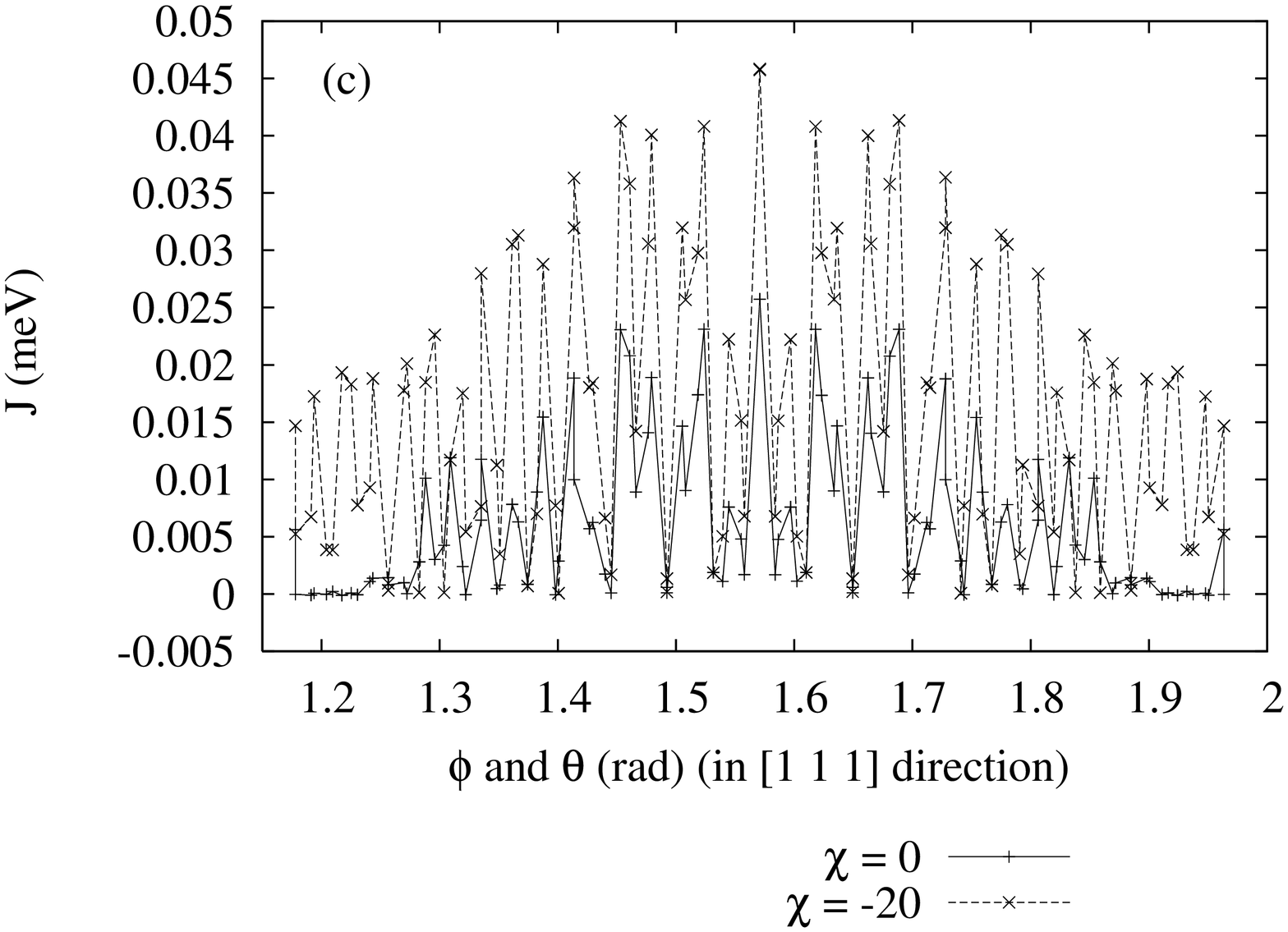} 
\end{center}
 \caption{\label{fig:figh3} Plot of the H-M exchange coupling for $|R|=14$nm and $\chi = 0$ and -20. In (a) we calculate $J(\bm{R})$ in the $xy$-plane, where $3 \pi /8 \leq \theta \leq 5 \pi /8$. In (b) we calculate $J(\bm{R})$ in the $yz$-plane, where $3 \pi /8 \leq \phi \leq 5 \pi /8$. In (c) we calculate $J(\bm{R})$ in the $[1 1 1]$-plane, where $3 \pi /8 \leq \theta, \phi \leq 5 \pi /8$. Here the marked points correspond to actual data points evaluated using the H-M method, the lines drawn are a guide for the reader only. }
\end{figure}

The results presented earlier in this section show the exchange coupling for varying magnitudes of $\bm{R}$ only along the $y$ axis. When the inter donor separation contains non-zero $R_x$, $R_y$ and $R_z$ terms, we see a marked difference from the relative smoothness of the exchange coupling curves with only a non-zero $R_y$ term. 

We show the exchange coupling for $|\bm{R}|=14$nm and $\chi = 0$ and $-20$,  for varying $\theta$ and $\phi$ in Fig.~\ref{fig:figh3}. The inter valley interference causing the wild oscillations in $J(\bm{R})$ is highest when the inter donor separation contains all non-zero $R_x$, $R_y$ and $R_z$ terms (i.e. in (c) where both $\theta$ and $\phi$ are varied). Similar results have already been reported\cite{cam,koiller} using H-L theory, and we have confirmed these results using our more extensive H-M basis.

For $\chi = -20$ we see that the exchange coupling changes dramatically for non-zero $R_z$ parts in Fig.~\ref{fig:figh3} (i.e. when $\phi$ varies). This is because for $\chi = -20$ the inter valley interference terms come only from the $\pm z$ valleys, as the single donor ground state orbitals favour the $F_{\pm z}$ valleys.\cite{koiller}

If we study the exchange coupling closely we can identify where the peaks and troughs occur in the exchange coupling as a function of $\theta$ or $\phi$. For example, if we examine Fig.~\ref{fig:figh3}(a) we can identify where the peaks and troughs occur for $\chi =0$. In calculating the two-electron Hamiltonian matrix and overlap matrix elements, we find that several of the integrals involving $\Psi^{e_n}_{Q_2}(\bm{r-R})$ in the integrand, have a common factor of $e^{i (\bm{k_\mu - k_\nu}).\bm{R}}$, in the sum over $\bm{k_\mu}$ and $\bm{k_\nu}$.

In Fig.~\ref{fig:figh3}(a), we are varying $\bm{R}$ in the $xy$-plane, and consider $\bm{R} = (R_x,R_y, 0)$, where $R_x = 14 \cos\theta$ and $R_y = 14 \sin\theta$, and $3 \pi/8 \leq \theta \leq 5 \pi/8$. We recognise that for the range of $\theta$ we consider, $R_y$ does not vary as much as $R_x$, and $R_x$ ranges over both positive and negative values. Thus we found that $R_x$ is the most significant factor in determining the peaks and troughs for this plot. 

In the simplest case when $\theta = \pi/2$ and $R_x = 0$, and $\bm{R} = (0,14,0)$, there is a peak in the exchange coupling. We find that this is due to the fact that the real part of $e^{i (\bm{k_\mu - k_\nu}).\bm{R}} $ is the maximum value of 1, for 18 out of the 36 possible combinations of $\bm{k_\mu}$ and $\bm{k_\nu}$. Similarly we observe peaks in the exchange coupling when $R_x = \pm m a^0 / k$, where m is any integer, and $k=0.85$. If $\bm{k_\mu} = \pm \bm{k_x} = \pm (k,0,0)2 \pi / a^0$, we find the real part of $e^{\pm i \bm{k_x}.\bm{R}}$ equals 1. Again we find that the real part of $e^{i (\bm{k_\mu - k_\nu}).\bm{R}} $ is 1 for 18 out of the 36 possible combinations of $\bm{k_\mu}$ and $\bm{k_\nu}$.

It is difficult to determine the magnitude of the peaks in the exchange coupling, because this magnitude also depends on terms involving $e^{\pm i \bm{k_y}.\bm{R}} = e^{\pm i (2 \pi k R_y)/a^0}$ which is a complicated function of $R_y$ and $\theta$. However in general we observe local maxima at values of $\theta$ for which $R_x = \pm m a^0/k$, where m is any integer. 

Conversely we find troughs in the exchange coupling occurring at values of $\theta$ for which $R_x = \pm m a^0 / (2 k)$, where m is an odd integer. Here we find that the real part of $e^{i (\bm{k_\mu - k_\nu}).\bm{R}}$ is 1 for 10 out of the 36 possible combinations of $\bm{k_\mu}$ and $\bm{k_\nu}$, and $-1$ for 8 of the remaining combinations. Thus at these values of $\theta$ we observe local minima in the exchange coupling. Because we are only able to evaluate the exchange coupling at finite grid points, the peaks and troughs were best matched to the data points available, thus enabling us to identify these trends.

For $\chi=-20$ in Fig.~\ref{fig:figh3}(a) we find that the exchange coupling is relatively constant. This is because for $\chi=-20$ the dominant contributions in the ground state come from the two $\pm \bm{k_z}$ valleys, and only very small contributions from the other four valleys. Thus when $\bm{R} = (R_x, R_y, 0)$, the real part of $e^{\pm i \bm{k_z}.\bm{R}}$ is 1 for all the combinations of  $\pm \bm{k_z}$, and since the ground state has its largest components only in the $\pm \bm{k_z}$ valleys, the exchange coupling is maximised and almost constant, for this orientation. The small fluctuations in the exchange coupling are most likely due to the fact there may be small contributions from the other four conduction band minima, which oscillate as a function of $\theta$ as we saw earlier. In (b) and (c) of this figure we observe that when $\bm{R}$ contains non-zero $R_z$ part, the exchange coupling may oscillate even more wildly for $\chi=-20$, as a result of the inter valley interference from only the two conduction band minima, $\pm \bm{k_z}$.

\section{Conclusions and future directions}

It is imperative to know precisely, the single electron wave function and two-electron states to determine accurately the parameter regime necessary for a nuclear spin or electron spin quantum computer. In this paper we provide detailed electronic structure calculations using a molecular orbital method, for a pair of P donor electrons in relaxed and uniaxially strained Si.

We have determined the excitation spectrum of two electrons on P donors in relaxed and strained Si, and studied its dependence on donor positioning in the Si lattice. In particular, we concentrated on the targeted ground state singlet and triplet ``H-L'' orbitals, and examined the isolation of these ground states, from the rest of the excited Hilbert space. Furthermore we calculated the exchange coupling and double occupancy probability as a function of strain and donor position. 

Both the exchange splitting and double occupancy probability have a similar dependence on inter donor distance and lattice positioning of the P atoms. Thus, a compromise is needed to maintain a very small double occupancy probability in zero field, while realizing a sizable exchange coupling during a gating action. In unison with previous theoretical studies of the exchange coupling of P donors in relaxed and strained Si,\cite{cam,koiller,koiller2} we found that the exchange coupling, (and thus double occupancy probability), is extremely sensitive to the relative orientation of the two P donors in the Si lattice. However, the energy level spectrum appears not to be affected by the relative orientation of the P donors, as these energies are on a much larger scale than the exchange coupling.

The oscillations in the exchange coupling due to inter valley interference, have serious implications for any quantum computer architecture that relies on the exchange interaction to couple qubits. This sensitivity can be reduced in the presence of strain, for displacements of the donors within the plane perpendicular to the direction of the uniaxial strain. We have also identified the values of $\bm{R}$ that lead to the peaks and troughs in the exchange coupling.

It would be useful to investigate the effect of an applied voltage on these oscillations, to examine the exchange coupling during and after a gate operation, to determine what device parameters are required for fault-tolerant quantum computation. It is also important to study if the gating action can be performed adiabatically, i.e. if during the evolution of the two-electron system there remains a finite gap between the ground and excited states.\cite{hu2}

Wellard \emph{et al.}\cite{cam2} have extended these calculations to include the voltage dependence of the exchange coupling within the Heitler-London framework. We hope to develop their results further by using our extended basis to calculate the voltage dependence of not only the exchange coupling, but also the energy spectrum of the two-electron system, and double occupancy probability of the ground states. The energy spectrum informs us if during the gating action, the coupled donor system is well isolated, and the higher excited states can be safely neglected. In addition, the double occupancy probability gives us an estimate of the error rate. However, including the electric field is a very computationally intensive task, as it requires evaluating the basis functions on our grid, over a much greater range of device parameters, which is by far the most time consuming part for our calculations.

The results presented here using effective mass theory provide a solid foundation for future device modelling. Ongoing work on this project is focusing on extending the multi valley calculations to investigate the effect of gate voltage on the oscillations in the exchange coupling. One would expect that as the gate voltage is turned on, it will become more favorable for the donor wave function to distort toward the gate. As a result, the donor wave function no longer has equal contributions from all six valleys, and the oscillations may smooth out as the inter valley interference effects decrease. We will need to implement the full molecular orbital approach to obtain the higher excited two-electron states accurately. Thus using this method we gain insight not only into the exchange coupling and double occupancy probability, but also the conditions required to perform adiabatic gate operations. To make these calculations more tractable, we envisage that the Hund-Mulliken basis should suffice in calculating the exchange coupling over a greater range of device parameters, as we have shown here that this is the best compromise between accuracy and speed.

\begin{acknowledgments}
L.M.K. and H.S.G. would like to thank G.J. Milburn, C.J. Wellard and L.C.L. Hollenberg for valuable discussions relating to this work. L.M.K. would like to thank the Centre for Quantum Computer Technology and the Centre for Computational Molecular Science at the University of Queensland, where most of the computational work was performed. We acknowledge supercomputing time from the following facilities: the Australian Partnership for Advanced Computing National Facility, University of Queensland Computational Molecular Science Cluster Facility, and the University of Queensland's high performance computers. L.M.K. and H.S.G. acknowledge support from the National Science Council, Taiwan (ROC), under grant numbers NCS94-2811-M-002-050 and NSC94-2112-M-002-028.
\end{acknowledgments}

\end{document}